\documentclass[5p]{elsarticle}

\usepackage{import}
\usepackage{amsmath}
\usepackage{amsfonts}
\usepackage{amssymb}
\usepackage{ragged2e}
\usepackage{multicol}
\usepackage{lscape}
\usepackage{caption}
\usepackage{xcolor}
\usepackage{soul}
\usepackage[utf8]{inputenc}
\usepackage[T1]{fontenc}
\usepackage{lmodern}
\usepackage{siunitx}
\usepackage{textcomp}
\usepackage{upgreek}
\usepackage{stfloats}
\usepackage{blindtext}
\usepackage{physics}
\numberwithin{equation}{section}
\usepackage{times}
\usepackage{setspace}
\usepackage{import}
\usepackage{enumitem}
\usepackage{etoolbox} 
\usepackage{lipsum} 
\usepackage{mathtools}
\usepackage{float}
\usepackage{booktabs}
\usepackage{tocloft}
\usepackage{pdflscape}
\usepackage{adjustbox}
\usepackage{rotating}
\usepackage{booktabs}
\usepackage{url}

\begin{document}
\begin{frontmatter}
\title{Bipolar plates for the next generation of proton exchange membrane fuel cells (PEMFCs): A review of the latest processing methods for unconventional flow channels}

\author[1]{Zahra Kazemi}

\ead{zahra.kazemi@mail.utoronto.ca}

\affiliation[1]{organization={Advanced Research Laboratory for Multifunctional Lightweight Structures (ARL-MLS), Department of Mechanical and Industrial Engineering, University of Toronto}, 
    city={Toronto},
    postcode={M5S 3G8}, 
    province={ON,},
    country={Canada}}

\author[1]{Kamran Behdinan}
\ead{kamran.behdinan@utoronto.ca}

\cortext[cor1]{Corresponding author}

\begin{abstract}
The rapid, unsustainable depletion of finite fossil fuel resources and their environmental consequences demand the deployment of affordable clean and sustainable energy solutions. Polymer electrolyte membrane fuel cell (PEMFC) technology is an important pathway in decarbonization of modern energy systems, especially when fueled by high-purity green hydrogen. In PEMFCs, bipolar plates largely determine cell efficiency, longevity, and affordability, which in turn depends on both material selection and design of the embedded flow channels. Conventional manufacturing processes have long been used to fabricate standard bipolar plate designs; however, they are incompatible with unconventional, intricate geometries due to their insufficient resolution and precision in fabrication of fine features, and reliance on multi-step post-processing modifications that limit their design adaptability. This lack of design flexibility impedes the translation of innovative laboratory-scale concepts to industrial-scale production and their practical adoption. In recent years, a growing body of research publications and patent disclosures has reported advanced manufacturing methods, such as additive manufacturing, capable of producing intricate bipolar plate geometries at competitive costs. However, a discussion of these manufacturing approaches, along with an assessment of their scalability and industrial readiness, remains absent in the literature. This study aims to fill this gap. It outlines recent progress and proposes future research directions toward affordable and efficient bipolar plate solutions for advanced PEMFC systems.
\end{abstract}

\begin{keyword}
Fuel cells \sep Clean energy \sep Bipolar plates \sep Unconventional flow routes \sep Advanced processing methods \sep Additive manufacturing
\end{keyword}
\end{frontmatter}



\begin{table}[ht!]
\centering
\footnotesize
\fbox{\begin{minipage}{\columnwidth}
\begin{tabular}{ll}
\textbf{Abbreviations} &  \\
ABS: & Acrylonitrile butadiene styrene \\
AFC: & Alkaline fuel cell \\
AM: & Additive manufacturing \\
BEV: & Battery electric vehicle \\
BJ: & Binder jetting \\
BP: & Bipolar plate \\
CL: & Catalyst layer \\
DLP: & Digital light processing \\
DMFC: & Direct methanol fuel cell \\
DMLS: & Direct metal laser sintering \\
DOE: & Department of Energy \\
EBF: & Electron beam free-form fabrication \\
EBM: & Electron beam melting \\
EIS: & Electrochemical impedance spectroscopy \\
EDS: & Energy dispersive spectrometry \\
FCEV: & Fuel cell electric vehicle \\
FFF: & Fused filament fabrication \\
GC: & Gas channels \\
GDL: & Gas diffusion layer \\
HOR: & Hydrogen oxidation reaction \\
ICE: & Internal combustion engine \\
LENS: & Laser-engineered net shaping \\
LMD: & Laser metal deposition \\
LPBF: & Laser powder bed fusion \\
MCFC: & Molten carbonate fuel cell \\
MEA: & Membrane electrode assembly \\
MJ: & Material jetting \\
MPL: & Microporous layer \\
ORR: & Oxygen reduction reaction \\
PAFC: & Phosphoric acid fuel cell \\
PBF: & Powder bed fusion \\
PEEK: & Polyether ether ketone \\
PEI: & Polyetherimide \\
PEKK: & Polyether ketone ketone \\
PEM: & Proton-exchange membrane \\
PEMEC: & Proton exchange membrane electrolyzer cell \\
PEMFC: & Proton exchange membrane fuel cell \\
PEMWE: & Proton exchange membrane water electrolysis \\
PLA: & Polylactic acid \\
SEM: & Scanning electron microscopy \\
SHS: & Selective heat sintering \\
SLA: & Stereolithography \\
SLS: & Selective laser sintering \\
SLM: & Selective laser melting \\
SMWS: & Selective microwave sintering \\
SOFC: & Solid oxide fuel cell \\
WAAM: & Wire arc additive manufacturing \\
\end{tabular}
\end{minipage}}
\end{table}

\small
\section{Introduction}

Fossil fuel–based energy sources are the main contributors to the environmental crisis due to the substantial pollutants they emit into the air, water, and soil. These emissions disrupt climate patterns, acidify oceans, contaminate natural resources, and pose health risks \cite{ang2022comprehensive, carrette2001fuel}. With global population growth and associated energy demands, reliance on finite fossil resources cannot be sustained. Given their ecological footprint and rapid, unsustainable depletion, a worldwide transition toward renewable and eco-friendly energy solutions is imperative. Renewable energy flows endlessly from natural resources such as solar radiation, flowing water and ocean waves, wind, and the geothermal heat from the Earth crust; these sources are abundant, continuously replenished, and regarded as viable alternatives to conventional energy sources. However, their large-scale integration into power systems remains limited because of their irregular availability and dependence on climatic and seasonal conditions. Thanks to recent progress in energy storage and smart grid technologies, they have covered 29.8\% of global electricity production in 2022 \cite{ostergaard2022renewable, surarapu2016emerging, mekuye2024energy}. Hydrogen energy based on renewables is another way for energy production.

There are several processes available for hydrogen production \cite{dawood2020hydrogen}. At present, about 96\% of hydrogen production relies on fossil fuel-based processes, which is commonly referred to as gray hydrogen \cite{hassan2024recent}. Alternative production routes include biomass gasification and water electrolysis. Hydrogen produced by the former is often contaminated with tar and impurities, whereas water electrolysis, typically powered by renewables and nuclear power \cite{doe_hydrogenp, hassan2024recent}, produces high-purity hydrogen. This green hydrogen can be consumed in fuel cells to produce electrical energy through electrochemical reactions, and to contribute to the development of low-carbon energy systems \cite{sorensen2018hydrogen, wee2007applications, felseghi2019hydrogen, manoharan2019hydrogen}. The main types of fuel cells currently being developed are molten carbonate (MCFC) \cite{dicks2004molten}, alkaline (AFC) \cite{ferriday2021alkaline}, proton exchange membrane (PEMFC) \cite{tellez2021proton}, solid oxide (SOFC) \cite{gur2016comprehensive}, direct methanol (DMFC) \cite{qasem2024recent}, and phosphoric acid (PAFC) \cite{qasem2024recent}. Their technical characteristics are listed in Table~\ref{tab:1}. They are primarily classified by the electrolyte they use. This categorization dictates the electrochemical processes that occur within the cell, the catalysts, the operational temperature, and ultimately application domain. Among them, PEMFCs have been broadly adopted to supply power to residential and commercial buildings, and industrial sites, as portable power sources, and particularly for vehicular applications \cite{wee2007applications, hong2018recent, sando2009research}. 


\begin{table*}[ht!]
\centering
\caption{Overview of the technical features of the main types of fuel cells \cite{gur2016comprehensive, tellez2021proton, ferriday2021alkaline, qasem2024recent, dicks2004molten, doe_hydrogen, boldrin2019progress, bernay2002prospects, moore1999comparison, mclean2002assessment, rocha2024review}.}
\footnotesize
{\begin{tabular}{l l l l l}
\toprule
Cell type & Common electrolyte material & Mobile ion & Typical fuel & Oxidant \\
\midrule
SOFC & Ceramics (e.g., yttria-stabilized zirconia) & $\mathrm{O^-}$ & $\mathrm{H_2}$, $\mathrm{C_nH_m}$, $\mathrm{CO_2}$, $\mathrm{CO}$ & Air-supplied $\mathrm{O_2}$ \\
MCFC & Liquid solution of metal carbonates (e.g., lithium, sodium) & $\mathrm{CO_3^-}$ & $\mathrm{H_2}$, $\mathrm{C_nH_m}$, $\mathrm{CO_2}$, $\mathrm{CO}$ & Air-supplied $\mathrm{O_2}$ \\
PAFC & Phosphoric acid solution & $\mathrm{H^+}$ & High-purity $\mathrm{H_2}$ & Air-supplied $\mathrm{O_2}$ \\
DMFC & Solid polymer membrane & $\mathrm{H^+}$ & $\mathrm{CH_3OH}$ & Air-supplied $\mathrm{O_2}$ \\
PEMFC & Solid polymer poly-perfluorosulfonic acid or Nafion & $\mathrm{H^+}$ & High-purity $\mathrm{H_2}$ & Air-supplied $\mathrm{O_2}$ \\ 
AFC & Aqueous potassium hydroxide solution & $\mathrm{OH^-}$ & High-purity $\mathrm{H_2}$ & Air-supplied $\mathrm{O_2}$ \\
\toprule
Cell type & Oxidation at anode & Reduction at cathode & Temperature [$\mathrm{C^{\circ}}$] & Cell voltage [$\mathrm{V}$]\\
\midrule
SOFC & $\mathrm{H_2+O^{2-} \rightarrow H_2O + 2e^-}$ & $\mathrm{\frac{1}{2}O_2 + 2e^- \rightarrow O^{2-}}$ & 600–1000 & 0.8–1.0 \\
MCFC & $\mathrm{H_2 + CO_3^{2-} \rightarrow H_2O + CO_2 + 2e^-}$ & $\mathrm{\frac{1}{2}O_2 + CO_2 + 2e^- \rightarrow CO_3^{2-}}$ & 600–700 & 0.7–1.0 \\
PAFC & $\mathrm{H_2 \rightarrow 2H^+ + 2e^-}$ & $\mathrm{\frac{1}{2}O_2 + 2H^+ + 2e^- \rightarrow H_2O}$ & 150–200 & 1.1 \\
DMFC & $\mathrm{CH_3OH + H_2O \rightarrow CO_2 + 6H^+ + 6e^-}$ & $\mathrm{\frac{3}{2}O_2 + 6H^+ + 6e^- \rightarrow 3H_2O}$ & 60–200 & 0.2–0.4 \\
PEMFC& $\mathrm{H_2 \rightarrow 2H^+ + 2e^-}$ & $\mathrm{\frac{1}{2}O_2 + 2H^+ + 2e^- \rightarrow H_2O}$ & 50–100 & 1.1 \\ 
AFC  & $\mathrm{H_2 + 2OH^- \rightarrow 2H_2O + 2e^-}$ & $\mathrm{\frac{1}{2}O_2 + H_2O + 2e^- \rightarrow 2OH^-}$ & 90–100 & 1.0 \\
\bottomrule
\end{tabular}}
\label{tab:1}
\end{table*}

\begin{table*}[ht!]
\centering
\footnotesize
{\begin{tabular}{p{1.05cm} p{2.25cm} p{2.5cm} p{1.7cm} p{9.4cm}}
\toprule
Cell type & Power output [$\mathrm{kW}$] & Power density [$\mathrm{kW/l}$] & Efficiency [\%] & Applications\\
\midrule
SOFC & $<1-3000$ & 0.8–1.0 & 35-43 & Distributed generation in industrial sites and residential areas; auxiliary power units \\
MCFC & $<1-1000$ & 0.01–1.0 & 45-47 & Utility-scale power plants; auxiliary power; large distributed generation systems \\
PAFC & $50-1000$ & 0.01-1.0 & $>40$ & Transportation and distributed power generation\\
DMFC & $0.001-100$ & 0.2–1.0 & 40 & Electronic and other portable devices\\
PEMFC& $<1-250$ & 3.1-5.7 & 53-58 & Backup power for emergency services; small distributed generation; electric vehicles \\ 
AFC  & $10-100$ & 0.01-1.0 & 60 & Military applications, spacecraft, and portable power \\
\bottomrule
\end{tabular}}
\label{tab:0}
\end{table*}

PEMFCs have been specifically developed for transportation because of their compact and space-efficient designs, high energy densities of up to 300 $\mathrm{Wh/kg}$, refueling times of 5 to 10 minutes, and reliable operation at freezing conditions down to $-30$ $^\circ$C, aspects that outperform those of (Li-ion) battery electric vehicles (BEVs) \cite{hu2020battery, toyota2020mirai, groger2015electromobility}. So far, several major automakers have already launched their fuel cell electric vehicle (FCEVs) models \cite{ehsani2018modern, eudy2017fuel}. Honda introduced the first commercial FCEV, the FCX Clarity, in 2008 \cite{bravo2024historical}. Hyundai launched the Tucson FCEV in 2016, followed by the NEXO in 2019, with a 95 kW fuel cell stack, a power density of 3.1 kW/l, and an estimated cruising range of 380 miles. The power density of the second-generation Mirai, launched in late 2020 by Toyota, reached 4.4 kW/l with end plates and 5.1 kW/l without them, along with an extended driving range of around 402 miles \cite{bravo2024historical, jiao2021designing, toyota2020mirai}. The stack power density including end plates is expected to reach 6.0 $\mathrm{kW/l}$ within the next few years at current densities of 3 to 4 $\mathrm{A/cm^2}$ and cell voltages between 0.7 and 0.8 V \cite{jiao2021designing}. In addition to passenger vehicles, PEMFCs have also been adapted for heavy-duty transportation applications \cite{hua2014status, eudy2017fuel, wang2020materials}. Despite recent technological progress, FCEVs remain less competitive than conventional internal combustion engine (ICE) vehicles in terms of durability and manufacturing costs. Accordingly, the U.S. Department of Energy (DOE) defined benchmarks to direct ongoing research and development efforts \cite{DOE_USDRIVE_Roadmap}.

\begin{figure*}[ht!]
    \centering
    \includegraphics[width=13cm]{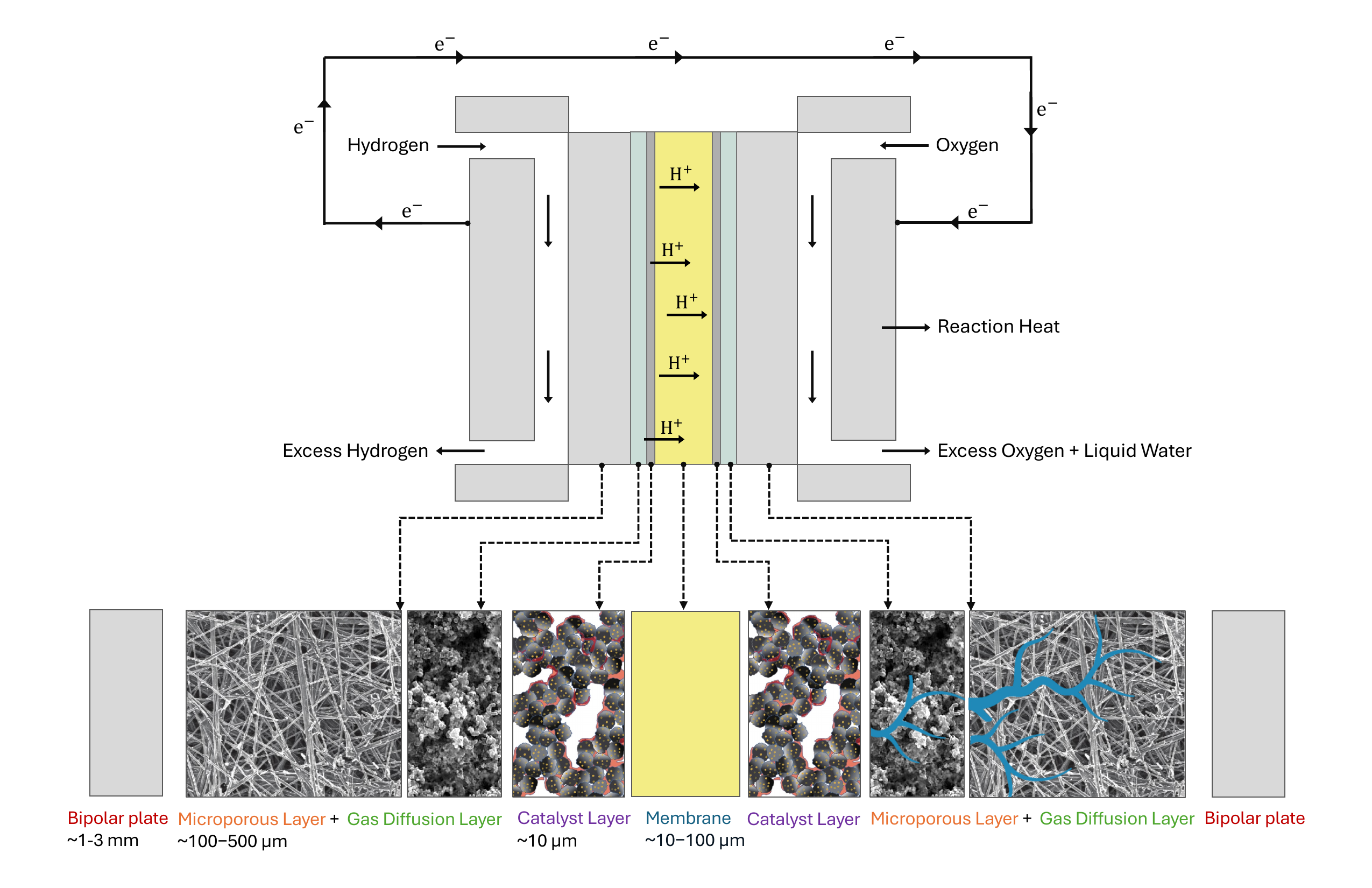}
    \caption{Schematic representation of a typical PEMFC stack and its operation principle. The upper illustration depicts the electrochemical processes where hydrogen at the anode loses electrons and oxygen at the cathode accepts electrons and protons to complete the reaction. The section shows a detailed cross-sectional view of the MEA and BP, morphology and characteristic thickness of each constituent layer. Data adapted with permission from \cite{kazemi2026topology}.}
    \label{fig:1}
\end{figure*}

A single PEMFC typically is a stack of bipolar plates (BPs) and a membrane electrode assembly (MEA), wherein there are gas diffusion layers (GDLs) coated with microporous layers (MPLs), catalyst layers (CLs), and a proton-exchange membrane (PEM) in the middle, as schematically illustrated in Figure~\ref{fig:1}. Through the hydrogen oxidation reaction (HOR) on the surface of the anodic CL, hydrogen molecules are split into protons and electrons. On the cathodic CL, oxygen molecules react with electrons and protons via the oxygen reduction reaction (ORR), and produce liquid water and heat. The performance of a PEMFC heavily depends on the multiphase transport behavior within its individual components, including the BPs, which in turn depends on the material and design of the embedded gas channels (GCs). Optimization of the transport properties of BPs is important for the proper functioning of the cell, as it (i) contributes to the homogeneity of reactant distribution over the catalyst sites and prevents the formation of local cold and hot spots; (ii) keeps a low pressure drop along the flow path and allows the PEMFC to operate without additional pumps or blowers for reactant recirculation; and (iii) supports the efficient drainage of byproduct liquid water to prevent flooding \cite{kumar2003effect}. From a material perspective, the BP must (iv) efficiently conduct electric current and dissipate reaction heat, (v) maintain chemical stability under the acidic and humid operating conditions, and (vi) must be gas-impermeable to prevent reactant leakage and gas crossover. (vii) Given that BPs make up 70-90\% of the total stack mass and volume \cite{wang2020materials}, a thin layer of material with low density-to-strength ratio is desirable. These multifunctional demands are reflected in the U.S. DOE targets for BPs \cite{DOE_USDRIVE_Roadmap, madheswaran2022polymer} and have motivated the majority of previous studies in the literature. 


BPs are conventionally made of graphite-based materials, either as graphite composites or as pure graphite \cite{song2020review}, because of their high conductivities and corrosion resistance. Nevertheless, graphite is brittle with poor processability, and its thin thicknesses needed for high power density are fragile \cite{yan2006performance, shimpalee2016investigation}. Graphite is also highly porous and gas-permeable; the recommended treatments to reduce its porosity introduce additional manufacturing steps and cost \cite{yan2006performance, shimpalee2016investigation}. Metal plates showcase superior mechanical behavior and are gas-impermeable. They are readily accessible with ease of manufacturing and high workability, which allows the fabrication of ultra-thin plates even below 1 mm \cite{zhang2024heat, song2020review, tawfik2007metal, fetohi2013ni, andre2010corrosion}. Metal plates, however, are prone to electrochemical corrosion and formation of passive metal oxide layer that thickens the plate and causes cell degradation \cite{karimi2012review, zhang2024heat, tawfik2007metal}. Protective surface treatments, such as noble-metal or metal-carbide coatings, are necessary for metal BPs \cite{zhang2011tin, wang2017ti}. Polymer composites reinforced with conductive carbon fillers have been also proposed. Many studies have developed composite BPs by inclusion of carbon additives (e.g., carbon nanotube and fiber, carbon black, graphite nanoplatelets) into polymeric matrices (e.g., polypropylene, polyphenylene sulfide, polyvinylidene fluoride, and phenolic resin). The fabrication routes for carbon-based and metallic BPs differ. Metallic BPs are typically manufactured using stamping and processes, whereas graphite-based plates are commonly produced by machining, compression molding, injection molding, or casting methods \cite{song2020review, kargar2020novel, hung2012fabrication, hu2014investigation}.

The mass transport capacity, defined by homogeneous reactant delivery to the catalyst surface at low pressure losses and efficient liquid water removal, is an important criterion for BPs and depends on the GC structure. Different channel profiles and layouts have been studied; rectangular, hemispherical, triangular, and trapezoidal profiles \cite{freire2014influence, yan2006numerical, liu2006reactant, kumar2003effect, wang2017respective, afshari2017investigation}, arranged in long serpentine paths, multiple short parallel paths, or interdigitated layouts are the most discussed \cite{manso2012influence, heidary2016experimental, gundlapalli2020performance}. In many cases, design parameters are often selected heuristically. Serpentine and interdigitated patterns distribute the reactant species evenly over the GDL. Serpentine channels minimizes the risk of water flooding, as liquid water is pushed out through a single flow path \cite{lim2016effects}, but they induce a notable pressure loss. Parallel configuration ensures minimal pressure drop due to shorter flow routes, but suffers from reactant maldistribution \cite{hontanon2000optimisation, lobato2010three} and water clogging at high current densities \cite{lim2016effects, gundlapalli2020performance}. Interdigitated designs generally outperform parallel patterns, but not serpentine designs \cite{spernjak2010situ}. These conventional BPs have been thoroughly discussed in the literature and adopted in practical PEMFC applications, largely because their acceptable, though not optimal, performance is balanced by low manufacturing cost and design simplicity. Despite their maturity, such conventional designs remain inadequate, as they typically depend on designer's limited intuition or heuristic modifications of previous designs with restricted performance. A stack power density of 6.0 $\mathrm{kW/l}$ or higher demands roughly a 20\% contribution from innovations in BPs \cite{jiao2021designing}.

\begin{figure*}[ht!]
    \centering
    \includegraphics[width=12cm]{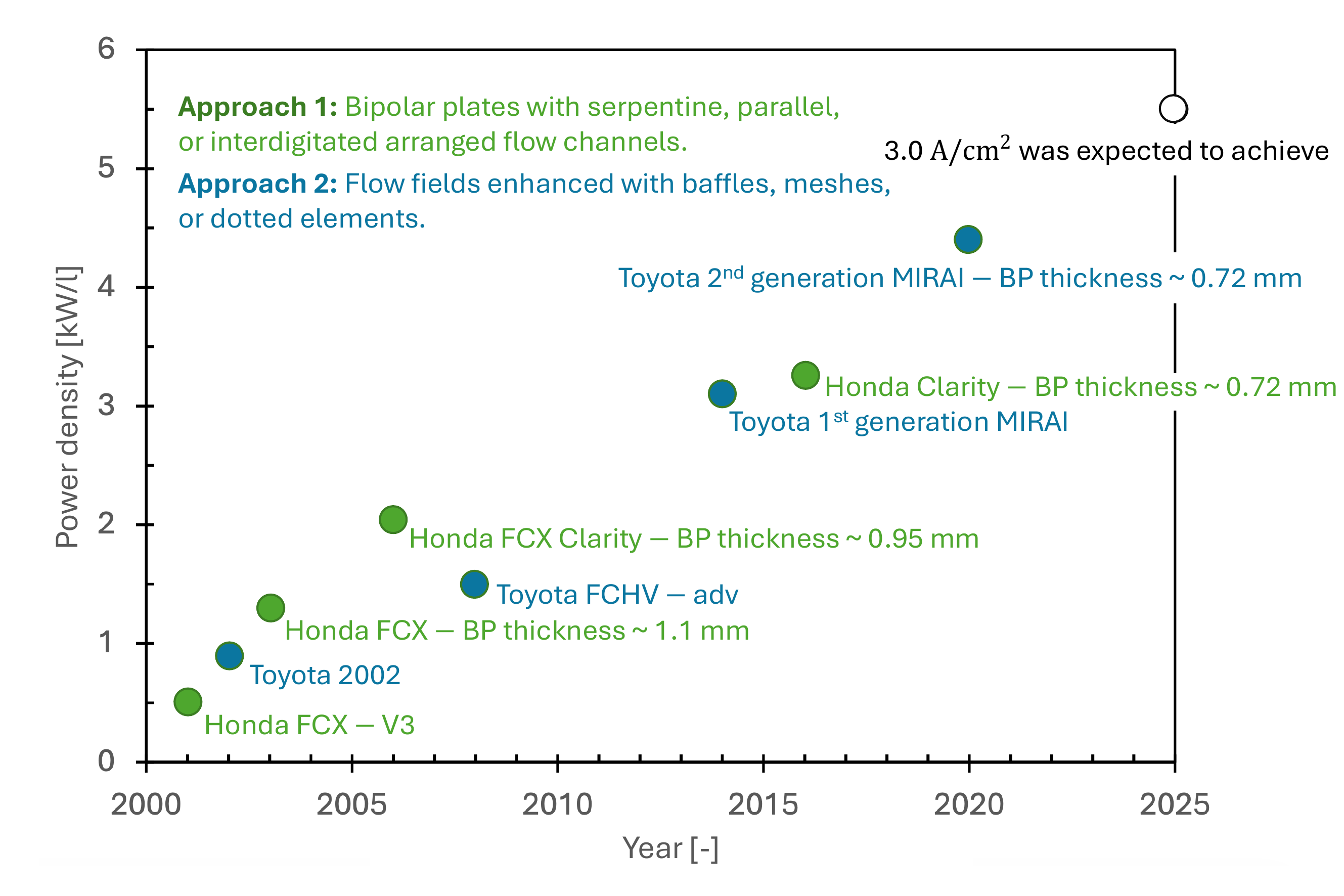}
    \caption{Progression of stack-level volumetric power density (including end plates) in Toyota and Honda FCEVs. The trend highlights how thinner bipolar plates and improved flow-field architectures have enabled more compact, high-performance fuel cell stacks over time \cite{jiao2021designing, konno2015development, saito2009new, inoue2005next, kikuchi2016development}.}
    \label{fig:4}
\end{figure*}

In the automotive industry, there have been two primary BP design routes; (i) incremental refinement of conventional layouts for narrower channel–rib structures, and (ii) the development of ribless BPs with microbaffles or microlattice structures (see Figure~\ref{fig:4}) \cite{konno2015development, saito2009new, inoue2005next, kikuchi2016development}. 

\begin{figure*}[ht!]
    \centering
    \includegraphics[width=16.25cm]{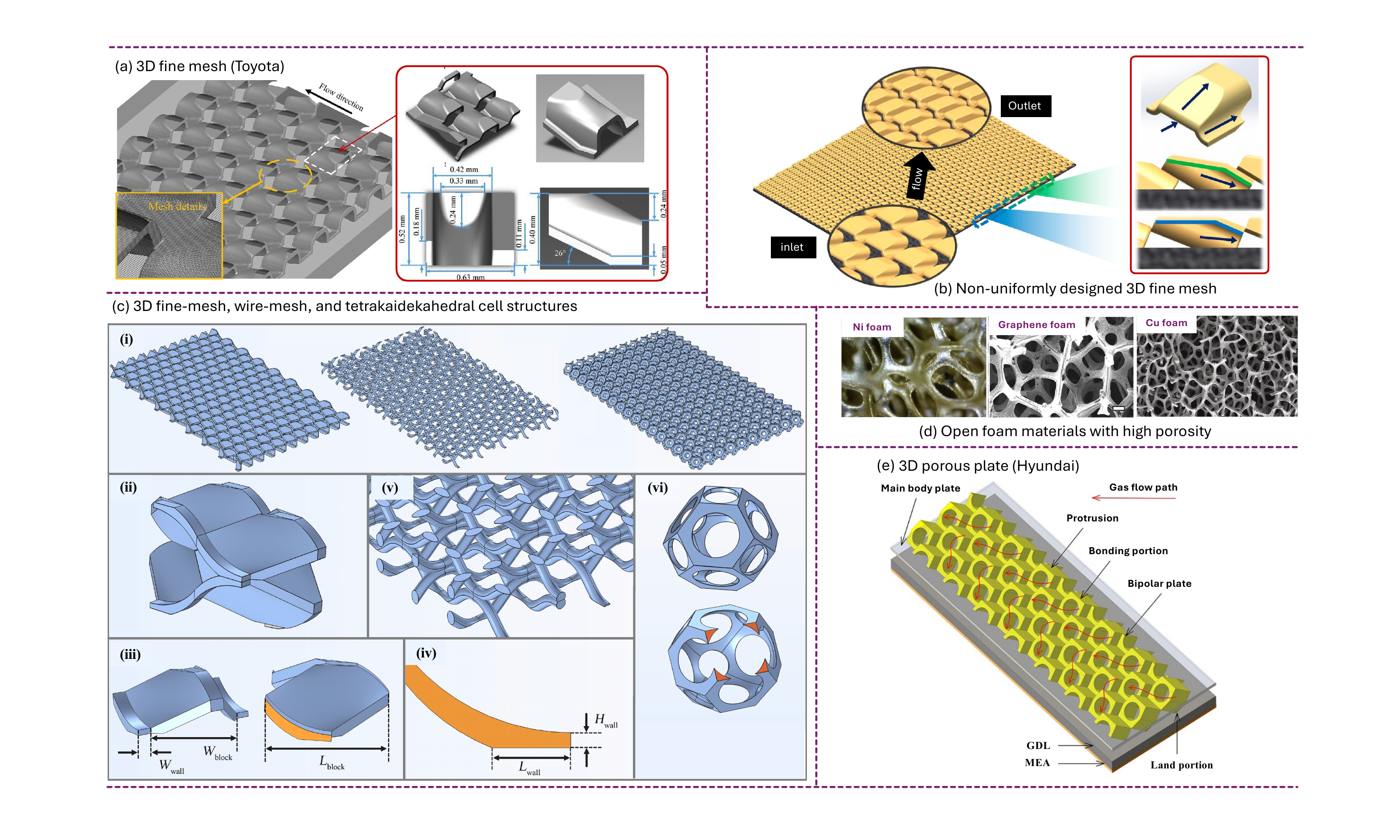}
    \caption{Examples of porous BP structures: (a) three-dimensional mesh BP adopted in the first generation Toyota Mirai, reproduced with permission from \cite{zhang2022porous, bao2019analysis}; (b) three-dimensional mesh with non-uniform arrangement of baffles, adapted with permission from \cite{choi2022experimental}; (c) representative fine mesh, wire mesh, and tetrakaidecahedron structures with localized enlargement and associated structural parameters, reproduced with permission from \cite{sun2024effects}; (d) Scanning electron microscopy (SEM) images of foam plates fabricated from Ni, graphene, and copper, reproduced with permission from \cite{shin2018effect, park2019gas, park2019achieving, sun2025porous}; and (e) porous BP used in the Hyundai NEXO, reproduced with permission from \cite{zhang2022porous}.}
    \label{fig:7}
\end{figure*}

\begin{figure*}[ht!]
    \centering
    \includegraphics[width=13cm]{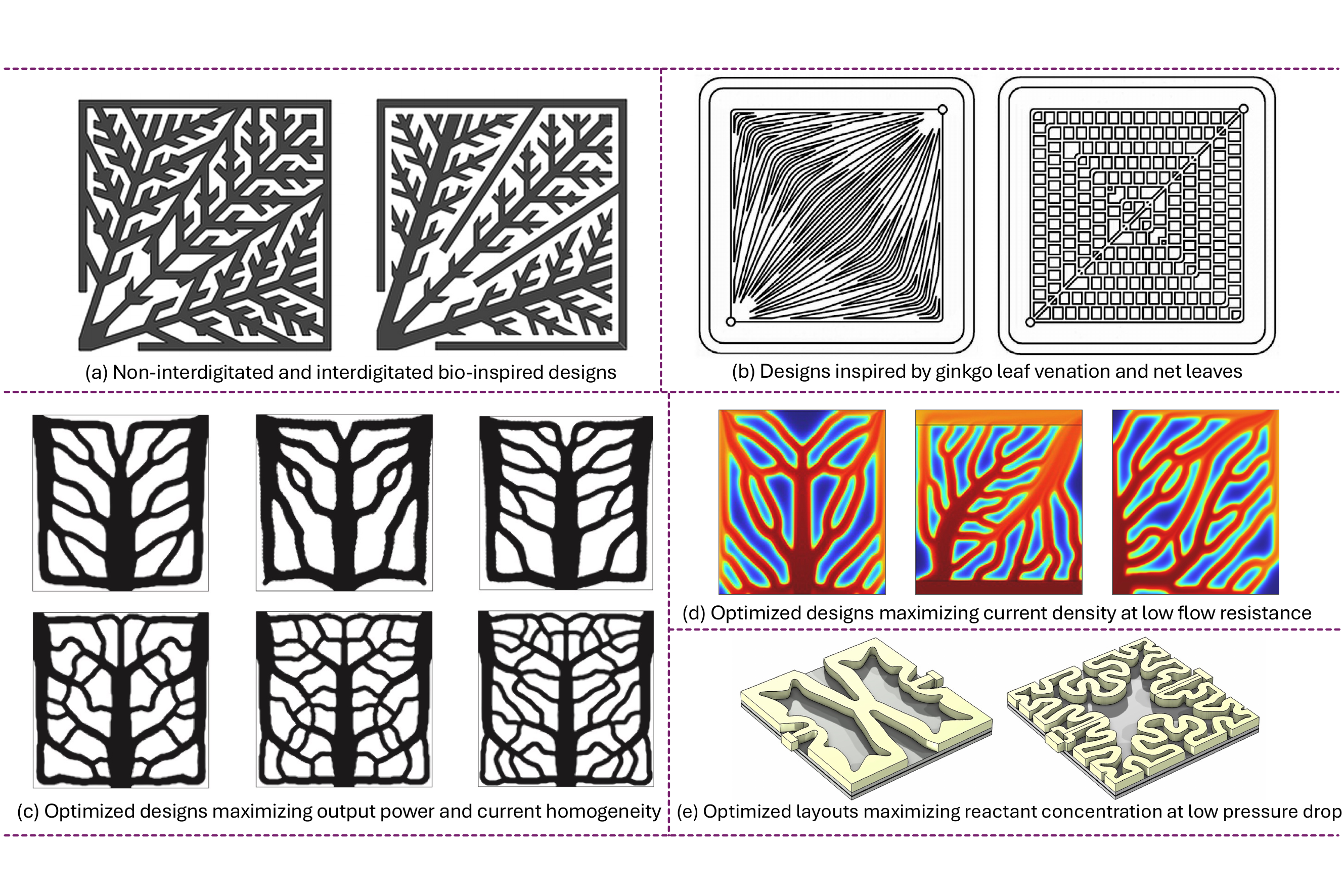}
    \caption{Examples of BPs with unconventional, non-intuitive flow patterns: (a) non-interdigitated, leaf-inspired configuration with constant channel width and interdigitated design with spatially varying channel widths determined by Murray’s law, adapted with permission from \cite{guo2014bio}; (b) gas channel layouts inspired by ginkgo leaf venation and net leaf patterns, reproduced with permission from \cite{kang2019performance}; (c) channel geometries obtained via reduced-order two-dimensional optimization to maximize output power and current density homogeneity over the current collector, reproduced with permission from \cite{behrou2019topology}; (d) flow geometries optimized to maximize mean current density while minimizing flow resistance for different inlet-outlet configurations, reproduced with permission from \cite{xia2024topology}; and (e) optimized layouts derived from three-dimensional and simplified two-dimensional topology optimization, formulated to maximize reactant concentration at the catalyst layer at low power dissipation, reproduced with permission from \cite{kazemi2026topology}.}
    \label{fig:8}
\end{figure*}

Microlattices integrated within flow channels, like those used in the first-generation MIRAI, are designed for heat dissipation and electric conduction inside the channels, and improved two-phase flow behaviour \cite{konno2015development, wang2009porous, park2019achieving, afshari2017investigation}. This integration allows the use of thinner BPs. Similarly, Shin et al. \cite{shin2018effect} found that metal foams inserted into BP channels as flow path (Figure~\ref{fig:7}d) increases the maximum output power by 50.6\% and shows more stable cell performance. They also recommended a dual-foam configuration with graded pore sizes that can boost the maximum power by up to 60.1\% \cite{shin2018effect}. Park et al. \cite{park2019gas} further proposed a unified BP-GDL into a highly porous graphene foam that functions as both (Figure~\ref{fig:7}d). This unified MEA reduced the cell thickness by 82\%, thereby lowers electron and mass transport resistances and showcases a higher volumetric power density compared to conventional MEAs. In addition to foam-based flow paths, three-dimensional baffled flow fields have been proposed. Periodically arranged small baffles accelerate reactant delivery toward the GDL as there is no rib-induced blockage, and when flooding intensifies along a flow route, water droplets escape into adjacent flow passages. An example of a uniformly patterned baffled plate with a constant baffle contact angle is shown in Figure~\ref{fig:7}a. Non-uniform baffle arrangements with varying baffle widths and contact angles (Figure~\ref{fig:7}b) also showcased a 12.8\% enhancement in power density at high current densities, largely because of reduced resistance in reactant and liquid water transport \cite{choi2022experimental, bao2019analysis}. Reactant maldistribution, commonly observed in conventional BP–GDL designs, has motivated the adoption of three-dimensional periodic lattice-based flow fields, such as fine-mesh and wire-mesh shown in Figure~\ref{fig:7}c. These highly ordered architectures ensure more even reactant delivery to the catalyst surface and hence, maximize catalyst utilization. Notably, Sun et al. \cite{sun2024effects} reported a 32.78\% increase in net power output for wire-mesh flow fields compared to conventional parallel-channel designs. Although mesh structures may locally reduce the electrical contact area at the MEA interface, the ohmic losses from current flow are, in most cases, negligible due to the small pore sizes and shortened electron conduction pathways. Biomimetic patterns inspired by natural flow distribution systems, such as those in lungs and leaf venation (Figure~\ref{fig:8}a and b), with enhanced multiphase transport capabilities are a class of of unconventional BPs with rib-channel style \cite{guo2014bio, kang2019performance}. In parallel, physics-based optimization approaches, such as shape optimization, gradient-based methods, evolutionary algorithms, and topology optimization with a large number of design degrees of freedom, have enabled the creation of unconventional, nonintuitive geometries tailored to maximize chosen performance metrics. For example, Figure~\ref{fig:8}c shows flow channels designed to boost power density and homogeneity of current distribution over the current collector \cite{behrou2019topology}. In Figure~\ref{fig:8}d, the optimized geometries improve electrochemical performance at low flow resistance \cite{xia2024topology}. Figure~\ref{fig:8}e showcases layouts obtained from three-dimensional and simplified two-dimensional topology optimization, formulated to maximize reactant concentration within the catalyst layer while keeping power losses along the flow paths to minimum \cite{kazemi2026topology}.



The realization and practical adoption of these unconventional yet functional BP design solutions depend on availability of advanced manufacturing routes that are both viable and industrially scalable. In recent years, several academic publications and patent disclosures has reported additive manufacturing methods capable of producing such intricate geometries at competitive costs. A discussion of these coupled design-manufacturing solutions with the latest contributions is therefore necessary to clarify research trend and guide further developments. This is the objective of the present review, a perspective that is largely absent from prior reviews as they typically consider these aspects in isolation or predominantly cover the conventional solutions.



\section{Additive manufacturing of PEMFC bipolar plates}

\begin{figure*}[ht!]
    \centering
    \includegraphics[width=17cm]{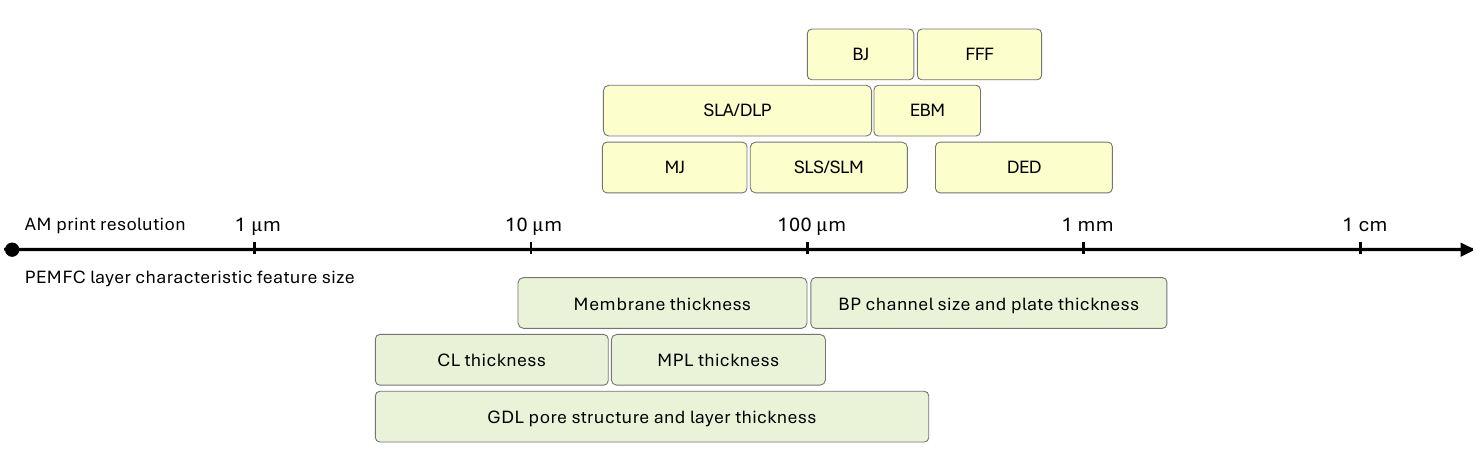}
    \caption{Comparison of characteristic feature sizes of PEMFC components with the achievable in-plane print resolution of major additive manufacturing methods. The logarithmic length scale (1 $\mathrm{\upmu m}$–10 cm) illustrates the dimensional hierarchy spanning catalyst layer thickness, microporous layer thickness, membrane thickness, gas diffusion layer microstructure, and BP channel geometry and thickness. The comparison highlights the extent to which different processes can fabricate PEMFC components across multiple length scales.}
    \label{fig:am}
\end{figure*}

Conventional manufacturing methods, such as molding, stamping, machining, pressing, casting have long been used in the production of conventional fuel cells \cite{mehta2003review}. However, the next generation of fuel cells that demand multifunctional layer designs with intricate, often nonintuitive geometries and micro- to nanoscale dimensions are largely incompatible with these conventional high-volume manufacturing processes. This is primarily due to their insufficient resolution and precision in fabrication of fine features, as well as reliance on multistep post-processing modifications that limit their design adaptability. This lack of design flexibility impedes the development and translation of innovative fuel cell concepts from laboratory demonstrations to industrial-scale production. In contrast, additive manufacturing with its layer-by-layer fabrication approach, allows the realization of unconventional, multiscale, and multimaterial PEMFC layers with high precision and a reduced number of production steps. A growing body of literature has studied the additive manufacturing of PEMFCs and has improved the technological maturity of PEMFC systems beyond what is achievable through conventional manufacturing routes. To date, various additive manufacturing processes, including powder-based processes, extrusion-based methods, photopolymerization processes, and jetting approaches, have been applied in the fabrication of PEMFCs \cite{astm2015astm52900}. These methods differ in both the feedstock form (e.g., liquids, filaments, pastes, or powders) and energy sources used for processing (e.g., including thermal heating, ultraviolet radiation, laser-based techniques, or electron-beam exposure) \cite{guo2013additive, ngo2018additive}. These choices influence achievable printing resolution and material deposition accuracy, microstructural control, and the applicability of each process for specific layers within the cell stack (see Table~\ref{tab:4}). For example, several studies have reported the successful fabrication of high-density metal-, polymer-, and graphite-based BPs, and metallic foams, with high dimensional accuracy and smooth surface finishes via electron beam melting (EBM) and selective laser sintering/melting (SLS/SLM) from the powder bed fusion family. Material jetting has been primarily used for catalyst layer deposition \cite{taylor2007inkjet, zubkova2026advanced} and, in some cases, for BP fabrication \cite{piri20203d}; however, its relatively low deposition rate makes it less favorable for large-area manufacturing. Vat photopolymerization, including stereolithography (SLA) and digital light processing (DLP), can reproduce extremely fine geometric details in BPs \cite{piri20203d, husaini2019digital, munoz2023engineering} and GDLs \cite{niblett2022utilization}, but are restricted to photopolymer resins. Material extrusion–based methods are compatible with a broader class of materials but typically showcase lower accuracy, process-induced defects, poor interlayer bonding, and pronounced mechanical anisotropy \cite{jang2022effects, yang2019novel, madheswaran2024mwcnt, madheswaran2025pemfc}. Figure~\ref{fig:am} maps the characteristic feature size requirements of PEMFC layers against the achievable in-plane print resolutions of major additive manufacturing processes, and shows the compatibility of each process with the multiscale structural demands of PEMFC fabrication. Additive manufacturing of PEMFCs is still at an early stage of development. To date, most reported studies have been limited to feasibility demonstrations and laboratory-scale implementations.

\begin{sidewaystable*}[p]
\centering
\scriptsize
\caption{Comparative overview of additive manufacturing processes applicable to PEMFC production, with emphasis on technological readiness and application-specific limitations \cite{taylor2007inkjet, zubkova2026advanced, piri20203d, husaini2019digital, munoz2023engineering, niblett2022utilization, jang2022effects, yang2019novel, madheswaran2024mwcnt, madheswaran2025pemfc}.}
\label{tab:4}
\begin{tabular}{p{2.7cm} p{1.3cm} p{3.8cm} p{3.8cm} p{11cm}}
\toprule
Category & Technology & Printable material & Printable layer & Description \\
\midrule
Vat photopolymerizatio & SLA & UV-curable photopolymer resins &  BPs (metallized post-printing), GDLs, self-breathing plates & \textbf{Merits}: Precision fabrication of unconventional intricate flow patterns with smooth surface finishes; excellent for rapid PEMFC prototyping. \textbf{Demerits}: Limited to polymeric material with relatively low mechanical strength and prone to deformation under clamping loads; electrically insulating (needs conductive coatings); limited chemical and thermal stability under acidic and humid PEMFC conditions. \\ \\
& DLP & UV-curable photopolymer resins & BPs (metallized via sputtering, electroless plating, etc.) & \textbf{Merits}: High in-plane print resolution for fine features with smooth surfaces. \textbf{Demerits}: Limited to polymer materials; electrically insulating; limited through-thickness resolution and anisotropy; prone to deformation and interfacial contact resistance; limited chemical and thermal durability under PEMFC conditions. \\ \\
\midrule
& SLS & Graphite or carbon powder blended with polymer binder (followed by debinding, carbonization, infiltration, sealing) & Graphite-composite BPs & 
\textbf{Merits}: Rapid fabrication of complex graphite geometries; moderate print resolution for typical dimensions; tunable mechanical strength, gas tightness, and conductivity via composition and infiltration. \textbf{Demerits}: Resolution limited by powder size and laser spot diameter; residual porosity and reactant leakage; multi-step post-processing; rough surfaces and higher contact resistance; lower conductivity than dense graphite or metals. \\ \\
Powder bed fusion & SLM & Metallic powders (316L stainless steel; Ni- and Co-based alloys) & Metallic bipolar plates; metallic foam GDLs & \textbf{Merits}: Enables highly intricate geometries with high precision and fine feature definition; near-full density achievable through optimized processing.
\textbf{Demerits}: As-built surface roughness and partially melted particles, higher pressure drop and contact resistance; internal porosity or lack-of-fusion defects, reduced gas tightness; residual stresses from thermal gradients; stainless steel requiring conductive, corrosion-resistant coatings. \\ \\
& EBM & Metallic powders (primarily Ti-based alloys) & Metallic BPs & \textbf{Merits}: Reduce residual stresses compared with laser-based PBF; well suited for reactive alloys like titanium; good densification achievable. \textbf{Demerits}: High as-built surface roughness requiring post-processing; corrosion and interfacial contact resistance issues persist; limited material selection and lower resolution than SLM. \\ \\
\midrule
Material extrusion & FFF & Thermoplastics (PLA, ABS, PEEK); conductive polymer composites or carbon-filled filaments & Polymer-based BPs (metal-coated for conductivity) & \textbf{Merits}: Low-cost, accessible method for rapid fabrication of BPs \textbf{Demerits}: Lower print resolution than SLA or DLP, with larger minimum feature sizes; pronounced layer roughness and interlayer voids; prone to gas leakage and increased interfacial contact resistance; strong anisotropy and dimensional inaccuracies; electrically insulating that require conductive coatings or metal lamination. \\ \\
& DIW & Catalyst inks (Pt/C + ionomer + solvent); conductive carbon-based inks or pastes & catalyst layers, patterned electrodes, and graded functional layers & \textbf{Merits}: Enables direct patterning of catalyst layers; compatible with highly viscous inks; efficient catalyst layer deposition with minimal material waste compared to spray coating methods. \textbf{Demerits}: Filamentary deposition can induce anisotropy and non-uniform distribution of Pt, carbon, and ionomer; nozzle clogging and shear-induced agglomeration are common; drying-induced stresses may lead to cracking and delamination; lateral resolution is limited and generally lower than that achievable with material jetting. \\ \\
\midrule
Binder jetting &  & Metallic or graphite-based powders with organic binder (followed by debinding, sintering, and infiltration) & BPs & \textbf{Merits}: Large-area fabrication with high precision, without high-energy beams, and minimal residual stress. \textbf{Demerits}: Post-processing (debinding, sintering, infiltration) required to achieve gas-tight, mechanical strength, and conductive plates; shrinkage, deformation, and dimensional inaccuracies possible during sintering; residual or interconnected porosity may cause gas crossover. \\ \\
\midrule
Material jetting &  & Catalyst inks (Pt/C + ionomer + solvents); ionomer dispersions for membranes (e.g., Nafion) & Patterned catalyst layers on membranes or GDLs; ultrathin membranes on catalyst layers & \textbf{Merits}: Precise control on local catalyst loading; ability to create compositional gradients; enables ultrathin membranes with strong electrode–membrane interface adhesion.
\textbf{Demerits}: Limited catalyst loading per pass and nozzle-clogging; droplet-based deposition may produce non-uniform layers or drying-induced cracks. \\ \\
\midrule
Directed energy deposition &  & Ni–Cr alloys, Ti-alloys, and other metallic feedstocks & Protective metallic coatings on BP substrates (surface modification) & \textbf{Merits}: Enables thick, dense, and metallurgically bonded coatings with excellent substrate adhesion for corrosion resistance and electrical conductivity; ideal for localized repair or refurbishment. \textbf{Demerits}:  High heat input and rapid deposition cause distortion, residual stress, and coarse microstructure; prone to gas leakage and interfacial contact resistance; often requires substantial post-processing. \\ \\
\bottomrule
\end{tabular}
\end{sidewaystable*}

\subsection{Fused filament fabrication}
Material extrusion is defined as process in which a semi-molten filament is fed through a nozzle and placed layer upon layer to build a structure. Fused filament fabrication (FFF), a primary subcategory of material extrusion, is one of the most accessible and adopted additive manufacturing methods. Thermoplastic polymers such as acrylonitrile butadiene styrene (ABS), Polyethylene Terephthalate Glycol (PETG), polylactic acid (PLA), and high-performance polymers including polyetherimide (PEI), polyether ketone ketone (PEKK), polyether ether ketone (PEEK), as well as carbon-reinforced polymer composites, are compatible with the FFF process \cite{kazemi2025uncertainty, sheikh2024fused, malekpour2026sustainable, azami2024additive, azami2025enhancing} for the rapid fabrication of unconventional geometries with fewer production steps. Optimal rheological behavior of the feedstock, with viscosity values below $10^4$ $\mathrm{Pa \,s}$ at low shear rates of $10^{-1}$ $\mathrm{1/s}$ and below $10^1$ $\mathrm{Pa \, s}$ at high shear rates of $10^2$ $\mathrm{1/s}$, ensures smooth printability and shape retention after placement \cite{acierno2023fused, solomon2021review}.

\begin{figure*}[ht!]
    \centering
    \includegraphics[width=16.5cm]{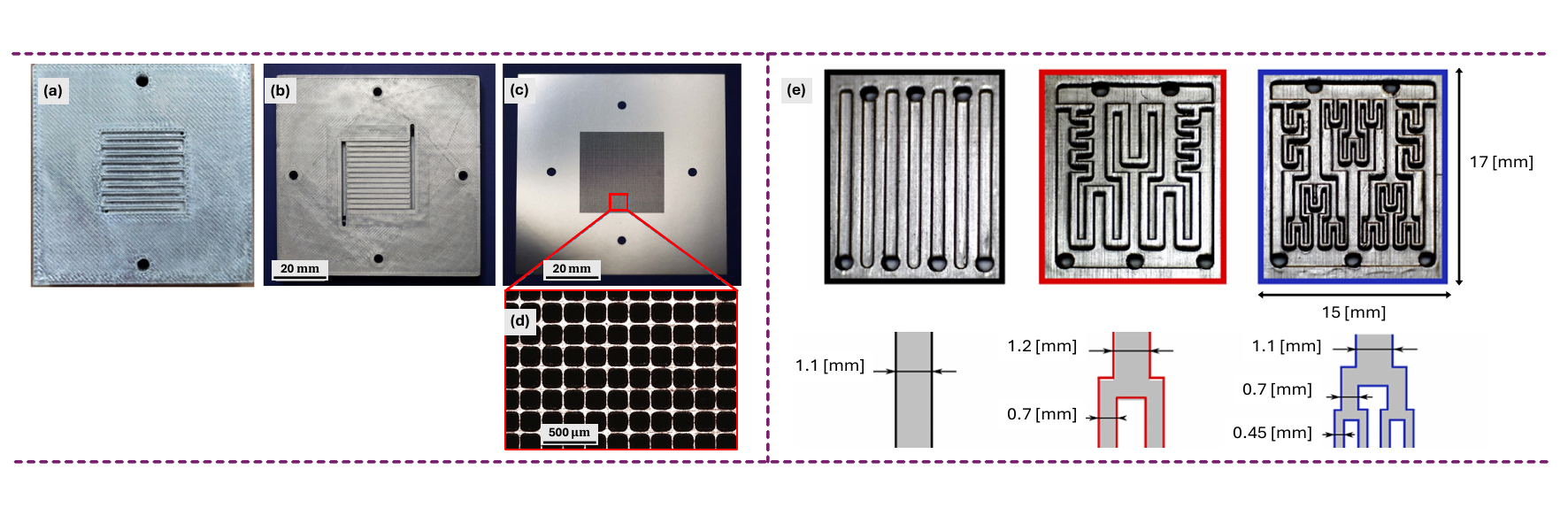}
    \caption{(a) Ag-coated FFF-printed BP with a serpentine flow field layout for PEMFC. Despite different thicknesses of the sputtered Ag current collector, the rough surface morphology and characteristic FFF printing patterns remain clearly observable, adapted under the Creative Commons Attribution 4.0 International License (CC BY 4.0) \cite{jang2022effects}. (b) and (c) show surface images of a PLA plate fabricated via the FFF process and an etched thin meshed titanium substrate at the GDL interface for enhanced electron conduction, respectively. (f) presents a microscopy image of the etched region, adapted with permission from \cite{yang2019novel}. Panel (e) shows macroscopic images of SLA-fabricated conventional interdigitated (as baseline) and lung-inspired flow-field plates coated with Ag, with channels geometric dimensions, adapted under the Creative Commons Attribution 4.0 International License (CC BY 4.0) \cite{munoz2023engineering}.}
    \label{fig:6}
\end{figure*}

Yang et al. \cite{yang2019novel} investigated the feasibility of using the FFF method to fabricate PEMFC BPs. They printed parallel channels, each $10 \times 1 \times 1$ mm, from PLA plastic (see Figure~\ref{fig:6}b). The printed plates were non-conductive and primarily designed for gas reactant supply and byproduct liquid water removal. To enable electron conduction, an ultra-thin titanium mesh, shown in Figure~\ref{fig:6}d, was placed at the GDL interface through a wet-etching process. The etched mesh, with a pore size of 260 $\mathrm{\upmu m}$ and a strut diameter of 17.8 $\mathrm{\upmu m}$, introduced a lateral electron conduction route that bypassed the BPs rather than following the conventional through-plane route. The study reported a production cost roughly one-tenth that of graphite plates and lower than that of electrically conductive metals, and a cell voltage of 2.21 V at 1 $\mathrm{A/cm^2}$. However, reliance on a conductive mesh and multiple post-print surface treatments to smooth print-induced roughness and visible layer lines added complexity to the fabrication process. The reported cell voltage also remained higher than that of conventional graphite or metallic BPs at comparable current densities and showed elevated ohmic overpotentials. High-frequency resistance measurements likewise exceeded those of conventional plates. These performance deficits reflected limited electrical conductivity at the PLA–titanium interface and less efficient charge transport, particularly at high current densities. Jang et al. \cite{jang2022effects} printed ABS polymer plates with a serpentine flow pattern, 0.8 mm deep and 1 mm wide. A thin Ag film was sputtered onto the plate surfaces for electron conductivity and to reduce the severe ohmic resistance caused by surface roughness. {\it{In situ}} characterization showed that, despite 8.22\% lower plate density, the open-circuit voltage reached over 0.9 V. Electrochemical impedance spectroscopy (EIS) measurements further showcased a maximum cell power density of 308 $\mathrm{mW/cm^2}$ with a 1.046 $\mathrm{\upmu m}$ thick Ag layer and a notable reduction in total ohmic resistance. However, the FFF-patterned surface remained clearly visible even after Ag deposition at different thicknesses, as shown in Figure~\ref{fig:6}a. The sputtered film mainly followed the underlying rough morphology rather than smoothing it. The noticeable geometric deviations and irregular channel walls could increase flow resistance, disturb boundary layer development, and alter water saturation behavior, which were not evaluated in the study. SEM observations also reportedly showed defects in the Ag coating, which might reduce long-term reliability. Furthermore, although the reported open-circuit voltage and peak power density were encouraging, a more accurate assessment of electrochemical performance would require comparisons with conventional BPs. Madheswaran et al. \cite{madheswaran2024mwcnt} manufactured composite BPs composed of a polyaniline matrix reinforced by multi-walled carbon nanotubes and patterned with serpentine flow channels 1.25 mm wide. At a nanotube loading of 25\%, the composite met the U.S. DOE targets with a mechanical strength of $40$ MPa, thermal conductivity of $20.29$ $\mathrm{W/mK}$, corrosion current density of 0.19 $\mathrm{\upmu A/cm^2}$ and corrosion protection efficiency of 97.29\%, and a cell output power density of 533.91 $\mathrm{mW/cm^2}$. Nevertheless, in-plane electrical conductivity remained suboptimal at 80.15 $\mathrm{S/cm}$. Despite post-processing treatments for surface smoothness, low surface quality and high interfacial contact resistance limited electrical performance. In a subsequent study \cite{madheswaran2025pemfc}, the authors substituted carbon nanotubes with graphene particles in the polyaniline matrix and observed comparable mechanical and electrochemical performance. However, poor electrical conductivity persisted, with in-plane and through-plane conductivities of 82.35 $\mathrm{S/cm}$ and 20.56 $\mathrm{S/cm}$, respectively, at a graphene loading of 10 wt\%.

The geometric freedom of the FFF process allows the creation of lightweight polymeric BPs with unconventional and non-intuitive designs. This is particularly important as they take up 70–80\% of the cell weight and 40–50\% of its manufacturing cost. However, the layer-wise fabrication and associated process-induced defects introduce multiscale inhomogeneities and cause inconsistencies in material properties compared to conventionally manufactured plates \cite{kazemi2025uncertainty, kazemi2022overall}. With a typical print resolution of $~$100 $\mathrm{\upmu m}$, FFF is less precise than many other additive methods for the fabrication of microscale details with high surface quality \cite{gao2021fused}. Although post-processing approaches, such as laser surface treatments, can reduce surface roughness and enhance the precision needed for micro-channel fabrication \cite{kim2021studies}, these additional steps increase manufacturing complexity and production time. Surface roughness and dimensional inaccuracies caused by warpage may disrupt reactant transport and induce leakage. The low electrical conductivity of polymers, combined with surface irregularities, further lead to inefficient charge transfer and suboptimal electrical performance. In FFF, the presence of interlayer voids and open porosity may further introduce the risk of reactant leakage.

\subsection{Vat photopolymerization}
Vat photopolymerization is a family of resin-based additive manufacturing processes wherein a UV laser or projected light source selectively cures a liquid photosensitive resin within a resin-filled vat. The most common variants in this category are SLA and DLP. SLA can be configured in either bottom-up or top-down setups and may also differ in its illumination approach. These methods are particularly distinguished by their high resolution, down to 20 $\mathrm{\upmu m}$, that allows for the fabrication of complex shapes, miscroscale features, and microporous structures with smooth surface finishes, high dimensional fidelity, and minimal post-processing treatments.

Piri et al. \cite{piri20203d}, in a feasibility study, printed cathode BPs with straight parallel flow channels of 4 mm $\times$ 0.4 mm square profile. The authors reported high-precision prints with a smooth surface finish and an average dimensional error of 2.09\% that shows the reproducibility of fine features below $0.5$ mm. This level of precision and size consistency, together with low manufacturing cost, makes SLA an ideal choice for the realization of functional BPs with delicate geometric features. The study mainly looked at the multiphase flow behavior in SLA-fabricated plates and analyzed the pressure and velocity profiles; however, aspects such as mechanical characteristics and electrochemical performance were not studied. Munoz et al. \cite{munoz2023engineering} also SLA-printed a lung-inspired design, coated it with conductive silver paint, and then sputtered it with platinum. The coated plates are shown in Figure~\ref{fig:6}e, with the sharp corners precisely reproduced. The printed plates were assembled into a cell stack and comprehensively tested during operation in terms of hydraulic behavior, pressure and velocity distributions, and electrochemical diagnostics. The SLA-printed lung-inspired designs homogenized the reactant distribution throughout the porous electrode with a higher mass transport rate to the reaction sites and outperformed the interdigitated design with higher output current densities at lower pressure loss. Husaini et al. \cite{husaini2019digital} fabricated coupons of bipolar plates via the DLP process. For electrical conductivity, the coupons were electrocoated with a nickel layer followed by a thin gold overlayer. At an electrocoating voltage of 5 V, the peak electrical conductivity reached around 400 $\mathrm{S/cm}$, with a low corrosion current density of 0.470 $\mathrm{\upmu A/cm^2}$, which met the U.S. DOE standards for bipolar plates. Porosity analysis confirmed that the coupons were dense and gas impermeable. This study was limited to {\it{ex situ}} material characterization, electrical conductivity measurements, and microstructural evaluation of porosity. {\it{In situ}} characterization within an operational cell and electrochemical behavior were not reported.

Compared with fused filament fabrication, interfacial contact resistance and altered flow behavior caused by surface roughness and dimensional deviations are less problematic in vat photopolymerization methods. Despite their potential, there has been a notable gap in the use of SLA and DLP in the manufacturing of PEMFC bipolar plates and they remain less common. This limited adoption is largely due to their incompatibility with metallic materials and high-performance thermoplastics such as PEEK, PEKK, and their composites, that can be processed via FFF. Vat photopolymerization is typically restricted to resin formulations with comparatively lower mechanical strength. Moreover, both FFF- and vat-printed materials often need additional modifications, such as inclusion of conductive fillers or surface coatings, to ensure adequate electrical conductivity and chemical stability.


\subsection{Powder bed fusion processes}
proofread this: Powder Bed Fusion (PBF) refers to a family of additive manufacturing processes where a thin, uniform layer of powdered material is selectively scanned by a focused energy source, whether a laser beam, electron flow, infrared lamp, or microwave pulses, to locally sinter or melt particles and bond them together. Major PBF variants include SLS, SLM, EBM, selective microwave sintering (SMWS), and selective heat sintering (SHS). SLS was originally developed to process polymer powders, particularly polyamides and polymer composites, and was later adapted for metallic materials through SLM and EBM. PBF processes typically use fine powder particles (often $5-25$ $\mathrm{\upmu m}$ and can create very thin layers and smooth as-built surfaces (surface roughness as low as $3$ $\mathrm{\upmu m}$ Ra). In contrast, extrusion-based methods rely on filament deposition that leaves visible layer lines and rougher surfaces. PBF is self-supporting, as the surrounding unfused powder supports the component during fabrication, thereby minimizes surface imperfections associated with support removal in extrusion methods. PBF can produce dense components with nearly isotropic mechanical properties comparable to those of conventionally manufactured materials. Thanks to these capabilities, PBF processes have been adopted to fabricate fine geometries in BPs with high dimensional fidelity and quality. Nevertheless, PBF processes demand controlled atmospheres, powder-handling procedures, and energy-intensive processing conditions. The steep thermal gradients and rapid solidification due to the high energy input can induce residual stresses, and post-processing heat treatments for stress relief. Although material extrusion and binder-based methods may offer lower costs and simpler manufacturing processes, they often sacrifice mechanical performance, material density, and dimensional accuracy.

\subsubsection{Laser powder bed fusion}
In laser powder bed fusion (LPBF) methods, laser energy is used to either fully melt or sinter powdered material \cite{azami2023laser}. LPBF is particularly recognized for its capability to process and manufacture high melting temperature materials, including ceramics and multiphase ceramic composites \cite{azami2023laser, azami2024comprehensive}. Careful optimization of LPBF process parameters, such as laser intensity, scanning velocity, and layer thickness, can improve print consistency, refine microstructure, and enhance material properties.

\begin{figure*}[ht!]
    \centering
    \includegraphics[width=16cm]{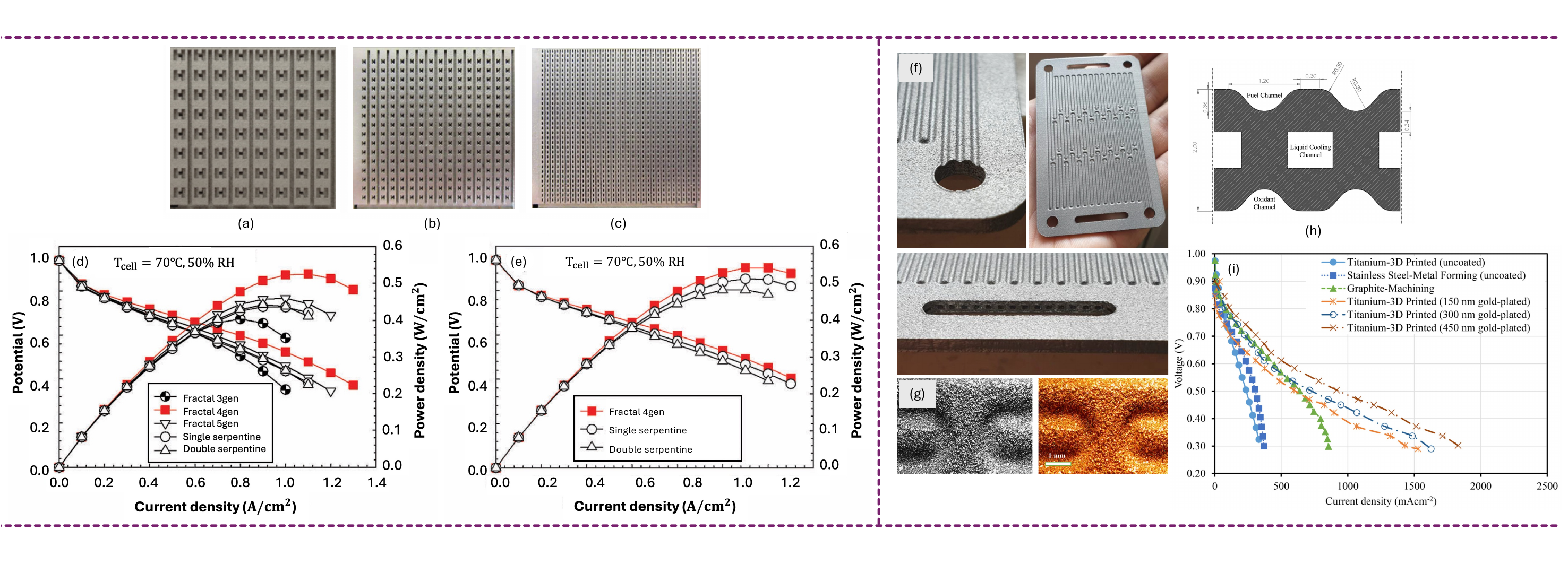}
    \caption{SLS-fabricated stainless steel BPs with a lung-inspired flow channel layout for three, four, and five levels of branching are shown in (a-c), respectively. The electrochemical performance of single cells using these plates is compared in (d) and (e) at 50\% relative humidity with active areas of 10 $\mathrm{cm^2}$ and 25 $\mathrm{cm^2}$. Data reproduced under the Creative Commons Attribution 3.0 Unported License (CC BY 3.0) \cite{trogadas2018lung}. (f-h) show an SLS-printed titanium flow plate with detailed flow field and cooling channel geometries (dimensions in mm), along with magnified surface morphologies in the as-printed condition and after 270 s of gold sputtering. The polarization curves in (i) compare the electrochemical performance of additively manufactured titanium BPs with conventionally manufactured graphite composite and stainless steel plates at 80$^\circ$C. Data adapted with permission from \cite{celik2022development}.}
    \label{fig:10}
\end{figure*}

The study by Chen et al. \cite{chen2004fabrication} was one of the first efforts to fabricate graphite-based BPs via selective laser sintering. In this study, a blend of a phenolic resin matrix and finely milled graphite was laser-sintered, then thermally converted to carbon and subsequently infiltrated with liquid epoxy to densify the as-printed material and reduce gas permeability. Bulk electric conductivity, flexural strength, corrosion resistance, gas leak and impermeability were characterized. Except for electrical conductivity, other properties evaluated met the prescribed DOE standards. A single cell was assembled with SLSed bipolar plates; however, {\it{in situ}} electrochemical performance testing was not reported. Dimensional inaccuracies due to shrinkage during carbonization and infiltration post-processing was a notable limitation. Bourell et al. \cite{bourell2011graphite} continued the earlier work by adding chopped carbon fibers to improve the mechanical strength of laser-sintered graphite bipolar plates. The addition of fibers enhanced the flexural strength from 35 MPa to nearly 50 MPa in finished parts; however, at a carbon fiber content of 26\%, the electrical conductivity decreased markedly from 400 to 50 $\mathrm{S/cm}$. This reduction was justified by the morphology and distribution of the constituent phases. At lower fiber loadings, however, the electrical conductivity could meet DOE specifications. Guo et al. \cite{guo2012effect} examined the mechanical strength and electrical performance of laser-sintered graphite composites composed of carbon black, synthesized and natural graphite, reinforced with carbon fibers, and bound with powdered phenolic. After carbonization, the porous brown materials were impregnated with a liquid epoxy resin and curing agent to become dense, gas-tight, and enhance mechanical properties. With optimized composition, the composites yielded flexural strengths between 30 and 50 MPa and electrical conductivities of $120-380$ $\mathrm{S/cm}$, values comparable to those of compression-molded and injection-molded graphite composite plates. However, lower densities were observed for laser-sintered composites. In {\it{in situ}} PEMFC testing, a single cell assembled with SLS-printed BPs yielded an electrical conductivity 120 $\mathrm{S/cm}$ and a flexural strength of 40 MPa. The tested BP had a serpentine flow field design with a thickness of 4 mm, channel width of 1.5 mm, and depth of 1.5 mm. Gould et al. \cite{gould2015performance} studied the direct metal laser sintering (DMLS) of Ti-6Al-4V bipolar plates. They printed plates of $40 \times 80 \times 3 \, \mathrm{mm^3}$ with triple-serpentine flow channels over an active area of $21 \, \mathrm{cm^2}$ for both a single cell and a 40-cell stack. Post-processing involved mechanical abrasion with catalytic surface activation to smooth and modify its chemistry, followed by acid etching to remove soluble impurities and deposition of $\mathrm{TiO_2}$ and gold coatings. The performance of the single-cell stack was comparable to that of a cell with conventionally manufactured bipolar plates; however, the 40-cell stack showcased 20\% lower than expected current density at 0.6 V. This reduction was primarily due to insufficient plate flatness, residual surface roughness, and conductive coatings, all of which contributed to high ohmic resistance within the stack. The plates exhibited an average flatness deviation of 44 $\mathrm{\upmu m}$, which became prohibitive when they were assembled in series. DMLS showed limited dimensional fidelity, with accuracy in the range of $100-200$ $\mathrm{\upmu m}$ and post-polishing surface height variations between $8$ and $99$ $\mathrm{\upmu m}$. Given that stack operation requires surface flatness below $3$ $\mathrm{\upmu m}$, this geometric limitation currently restricts DMLS-fabricated plates to single-cell studies or short stacks (fewer than five bipolar plates). Netwall et al. \cite{netwall2013decreasing} lessoned the interfacial contact resistance associated with the DMLS fabrication of titanium-alloy BPs for fuel cell assemblies. They proposed a compression break-in cycle designed to reduce the GDL–BP interfacial resistance. The lowest contact resistance reported was 4.75 $\mathrm{m\Omega/cm^2}$ at 1 MPa for a $\mathrm{TiO_2}$ coating overlaid with gold microdots on a titanium-alloy plate with a surface roughness of 1.6 $\mathrm{\upmu m}$. Although considerably reduced, this resistance still contributed appreciably to the overall ohmic losses of the cell. {\it{Ex situ}} test further showed that stack assembly compressed at $>1$ MPa eliminates GDL contact resistance. The authors also evaluated different BP topologies as the plate geometry affects contact resistance with the porous GDL. Guo et al. \cite{guo2014bio} fabricated graphite composite BPs with flow patterns inspired by plant leaves via the SLS process. Conventional parallel and interdigitated layouts were also printed for comparison. In all designs, the channel and land widths ranged from $1$ to $2.5$ mm. A peak power density above 0.5 $\mathrm{W/cm^2}$ was reported for the bio-inspired plates, which was $20-25\%$ higher than that of the conventional designs. This improvement was due to the more uniform reactant distribution and enhanced mass transport capabilities of the branched, leaf-like structure. Nevertheless, the surface finish of the SLSed plates appears relatively rough. Important parameters, such as geometric accuracy and surface roughness, as well as their effects on interfacial contact resistance and ohmic losses were not discussed. A detailed microstructural characterization, particularly regarding porosity, internal defects, and potential gas permeability, was not reported. The absence of a comparison with conventionally manufactured BPs was also a limitation of the study. Simialrly, Trogadas et al. \cite{trogadas2018lung} designed lung-inspired fractal flow fields with three, four, and five branching generations for homogeneous gas delivery across the catalyst layer in PEMFCs (see Figure~\ref{fig:10}a-c).s The flow plates were manufactured via DMLS of stainless steel, gold-plated, and assembled into the cathode of a PEMFC, with a conventional double-serpentine flow field mounted at the anode. In the five-generation design, the smallest channel profile measured $200 \times 300$ $\mathrm{\upmu m}$, with a manufacturing resolution of $33 \times 33$ $\mathrm{\upmu m}$. Radiographic imaging confirmed that the internal geometries were well-defined and defect-free. The four-generation design delivered a $20-25$\% increase in peak power density at current densities above 0.8 $\mathrm{A/cm^2}$ and reduced pressure losses by 50\% compared with conventional serpentine-based PEMFCs. In a study by Celik et al. \cite{celik2022development}, titanium BPs for PEMFCs were manufactured via SLS additive manufacturing. Graphite composite BPs were also manufactured using conventional machining, and stainless steel plates were manufactured through metal forming for comparison. All BPs had a width of 55.25 mm, a length of 92 mm, and a thickness of 2 mm. The detailed geometric dimensions of the reactant and internal cooling channels are presented in Figure~\ref{fig:10}h.The peak power density reached 639 $\mathrm{mW/cm^2}$ at a 450 nm gold coating on the titanium plates, which was twice that of the graphite composite plate and 3.7 times higher than that of the stainless steel plates. Excellent surface characterization results were reported for both coated and uncoated additively manufactured materials. Figure~\ref{fig:10}g shows magnified images of the as-printed surface and the gold layer thickness of 450 nm. Performance comparisons showed that graphite plates showcased a higher peak power density (322 $\mathrm{mW/cm^2}$) than uncoated titanium (119 $\mathrm{mW/cm^2}$) and stainless steel (173 $\mathrm{mW/cm^2}$). Once coated, the titanium plates outperformed the uncoated materials regardless of coating thickness, primarily due to enhanced electrical conductivity and reduced interfacial resistance. 


\begin{figure*}[t!]
    \centering
    \includegraphics[width=13cm]{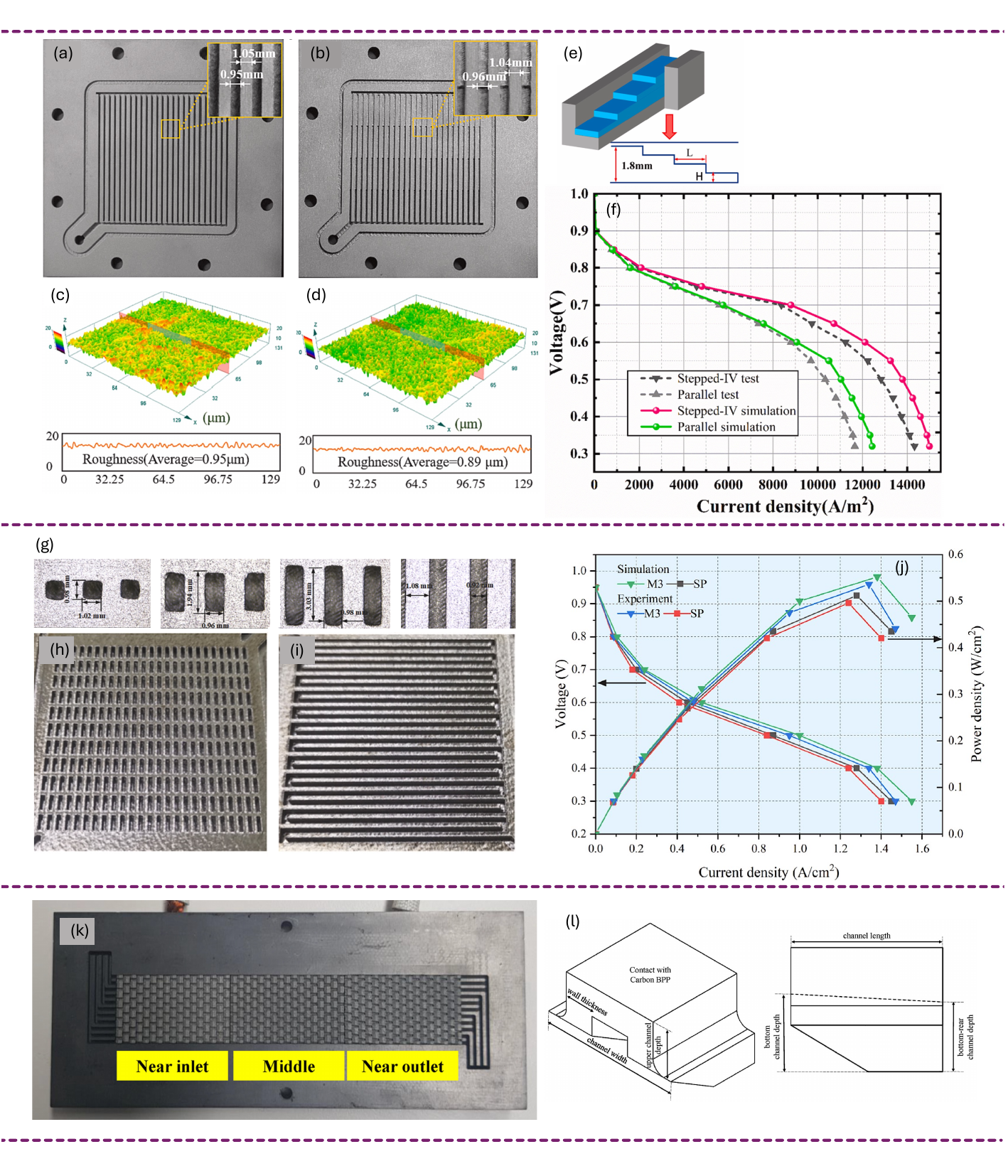}
    \caption{Surface morphology and characteristics of SLM-fabricated 316L stainless steel BPs are shown for both (a, c) simple and (b, d) stepped parallel flow field configurations. The additively manufactured plates had well-defined channel geometries and planar surfaces with a measured surface roughness of $0.89-0.95$ $\mathrm{\upmu m}$. Panel (f) compares the polarization behavior of a single PEMFC incorporating the printed parallel and stepped flow channels, as obtained from numerical simulations and experimental {{\it{in situ}}} measurements. Data adapted with permission from \cite{lu2024numerical}. Panels (g-i) present the macroscopic morphology and geometric dimensions of SLM-manufactured mesh and serpentine BPs after surface treatment. The polarization curves for these configurations are presented in (j). Data adapted from \cite{zhang2023investigation}. (k) shows a cathodic carbon BP with non-uniformly arranged laser-printed metallic baffles, and (l) illustrates the schematic geometry of an individual baffle. Data adapted with permission from \cite{choi2022experimental}.}
    \label{fig:13}
\end{figure*}

\begin{figure*}[ht!]
    \centering
    \includegraphics[width=14cm]{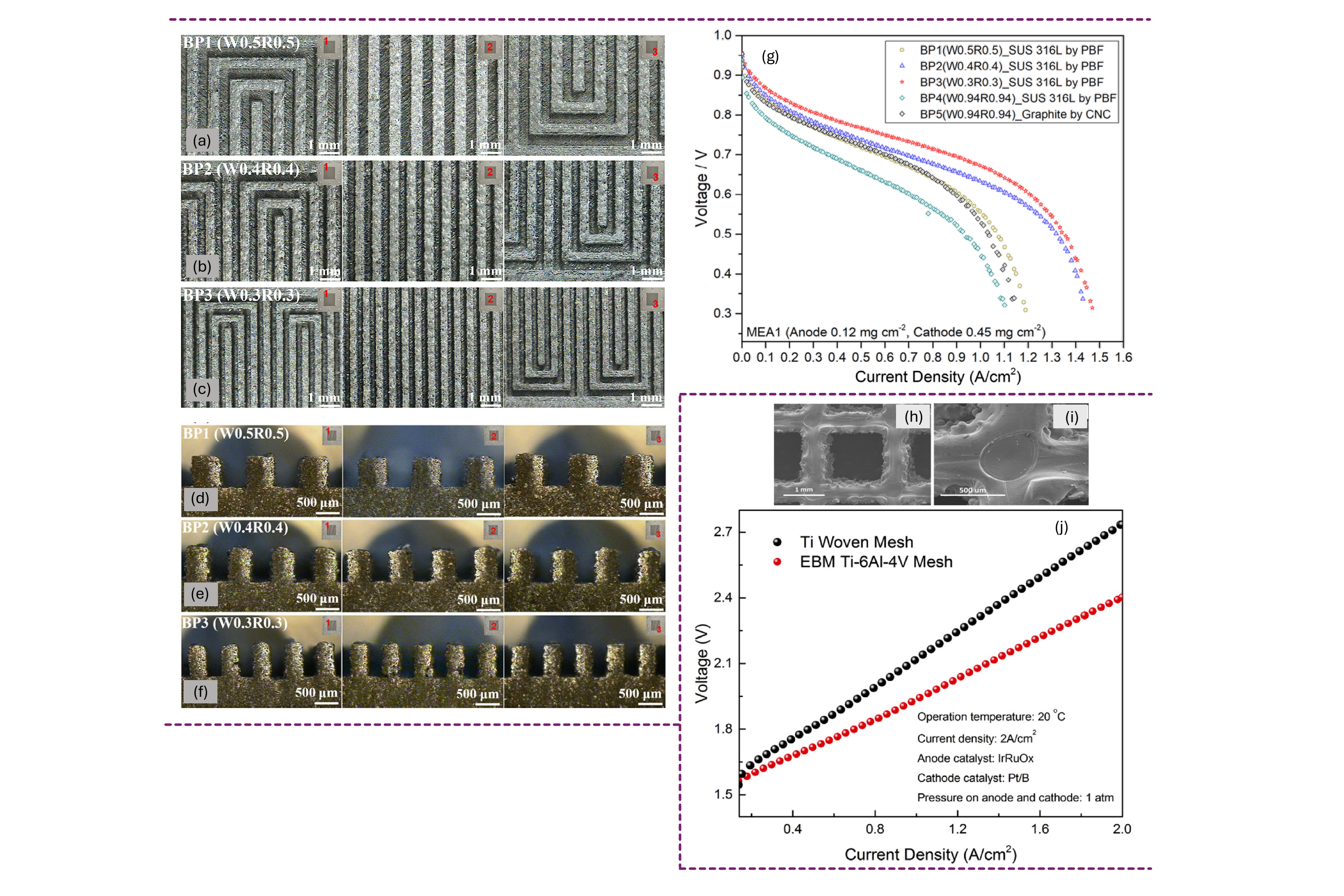}
    \caption{Top-view optical micrographs of laser-printed stainless steel BPs with a serpentine flow field design are presented in (a–c), corresponding to channel widths of 0.5 mm, 0.4 mm, and 0.3 mm, respectively, along with their cross-sectional profiles shown in (d-f). The $\mathrm{I-V}$ polarization performance of single cells are also compared across five different channel–rib sizes using a MEA containing Pt loadings of 0.12 $\mathrm{mg/cm^2}$ at the anode and 0.45 $\mathrm{mg/cm^2}$ at the cathode. Data adapted with permission from \cite{jin2024stainless}. SEM images of titanium alloy mesh GDLs fabricated via EBM are shown in (h) and (i). Panel (j) presents the polarization behavior of PEMECs using these mesh structures as anode GDLs, evaluated at room temperature. Data adapted with permission from \cite{mo2016additive}.}
    \label{fig:12}
\end{figure*}

In a preliminary study, Dawson et al. \cite{dawson2015investigation} manufactured metallic PEMFC BPs via SLM of 316L stainless steel. They printed a serpnetine design on the anode side and a simple parallel design for the cathode side, with a channel width of 1 mm. As expected, minor surface roughness and corrugation, associated to the powder particle size, were observed in the as-printed condition. After gentle surface polishing, the plates achieved appreciable contact with the MEA. The contact resistance and electrochemical performance of the SLM-fabricated plates were comparable to those of conventionally machined. Yang et al. \cite{yang2017additive} similarly additively manufactured 316L stainless steel BP for proton exchange membrane electrolyzer cells (PEMECs). The fabricated plate had dimensions of $60.7 \times 60.7 \times 5.4 , \mathrm{mm^3}$ and a parallel flow field design. In this study, the SLM-fabricated BP was, for the first time, tested {\it{in situ}} as the cathode BP in a PEMEC, paired with a conventionally machined graphite plate on the anode side. The additively manufactured BP was characterized by SEM and energy dispersive spectrometry (EDS), and a consistent elemental composition throughout the plate and slight oxidization were reported. The ohmic resistance was 0.229 $\mathrm{\Omega \, cm^2}$ at 20$^\circ$C and 0.129 $\mathrm{\Omega \, cm^2}$ at 80$^\circ$C. For a PEMEC with conventional graphite BPs on both sides, the ohmic resistance was reported to be 0.203 $\mathrm{\Omega \, cm^2}$, which was higher than that of the PEMEC with the SLMed plate. Overall, the PEMEC achieved 83.4\% efficiency at 1.779 V and 2.0 $\mathrm{A/cm^2}$. Lu et al. \cite{lu2024numerical} proposed a modified parallel flow field with stepped channels and manufactured stainless steel BPs via the SLM method. The SLM process enabled precise reproduction of channel geometries, with a channel width of 0.95 mm and an outlet depth of 1.8 mm, at an accuracy of 0.05 mm (see Figure~\ref{fig:13}a and b). The polarization test results presented in Figure~\ref{fig:13}f show that the stepped configuration performed much better in both ohmic and concentration regions, and achieved a peak power density of 6727 $\mathrm{W/m2}$ at 0.55 V, which was 26\% higher than the standard parallel layout. Moreover, the fabricated plates had well-defined geometries with precisely reproduced sharp corners and a low surface roughness of $0.89-0.95$ $\mathrm{\upmu m}$. Zhang et al. \cite{zhang2024heat} developed a novel baffled flow field by addition of sub-baffles at different inclination angles. Such intricate bipolar plates, with features as small as 0.3 mm for a 90$^\circ$ sub-baffle, were realized using SLM of stainless steel powder at a dimansional accuracy of 0.05 mm. Their performance was assessed solely through polarization measurements to validate the numerical models. In another study by Zhang et al. \cite{zhang2023investigation}, mesh and serpentine BPs were also fabricated through SLM method. The printed flow channels had dimensions on the order of $1-2$ mm (see Figure~\ref{fig:13}g-i). To reduce the interfacial contact resistance with the MEA, the as-printed plates were laser-treated and sandblasted. The dimensional inaccuracy of the mesh structure was reported to be less than 4\%, with an average surface roughness of 0.135 $\mathrm{\mu m}$ after surface treatment. The serpentine design showcased a larger dimensional deviation of about 8\% and a rougher surface finish. The SLMed plate demonstrated marginally better contact resistance and polarization behavior than the conventional serpentine design, with a maximum power density of 0.547 $\mathrm{W/cm^2}$ at 0.4 V, which was an 8.96\% higher than that of the serpentine design. This improvement resulted from the uniform transport of reactants within the mesh and the low contact resistance at the MEA interface. Choi et al. \cite{choi2022experimental} experimentally studied a non-uniform arrangement of tapered baffles with varying widths and sleep slope (Figure~\ref{fig:7}b). This geometry was specifically designed to lower mass transfer resistance, as there is no rib-induced blockage \cite{choi2022experimental, bao2019analysis}. In this design, the baffles gradually become narrower and steeper toward the outlet region, where flooding typically causes severe limitations on mass transport. The metallic baffles were fabricated via laser powered bed fusion and inserted into the cathode carbon BP of a PEMFC, as shown in Figure~\ref{fig:13}k. Due to manufacturing constraints, the channel wall thickness was restricted to 0.6 mm, and both the upper and lower channel depths were set to 1.0 mm. The $\mathrm{I-V}$ curve showed a 12.8\% increase in peak power density compared to parallel channel layout. However, the study did not provide information regarding the printing method or the material used, and the print quality was not discussed. Jin et al. \cite{jin2024stainless} also fabricated stainless steel BPs with serpentine micro-rectangular channels via the SLM method with high precision. Single fuel cells were then assembled with these laser-printed plates and a MEA containing low platinum (Pt) loadings of 0.12 $\mathrm{mg/cm^2}$ at the anode and 0.45 $\mathrm{mg/cm^2}$ at the cathode to investigate effect of channel widths on cell performance. The printed channels had high fidelity, with all channel corners accurately reproduced and no blockages observed, as seen in Figure~\ref{fig:12}a–c. Notably, surface roughness as low as 4.5 $\mathrm{\upmu m}$ was measured for the 0.3 mm-wide channels. The authors reported a maximum power density of 1.205 $\mathrm{A/cm^2}$ for the channel width of 300 $\mathrm{\mu m}$. This was 31.4\% higher than that of 500 $\mathrm{\mu m}$ channels, 70.2\% higher than 940 $\mathrm{\mu m}$ channels, and 34.9\% higher compared to a cell with CNC-machined graphite BPs of the same channel dimensions (see Figure~\ref{fig:12}g).


\subsubsection{Electron beam melting}
Electron beam powder bed fusion (EB-PBF), commonly known as electron beam melting (EBM), uses a focused flow of electrons to selectively melt powder within a high-vacuum chamber at high temperatures \cite{sadeghi2019influence}. Unlike LPBF processes, which are highly sensitive to the surface reflectivity and absorptivity of the feedstock material \cite{azami2023laser}, EBM is governed by electron-matter interactions. EBM is compatible with reactive metals and superalloys. For non-metallic materials and ceramics, conductive or metallic additives are often included to improve powder flowability and energy absorption within the melt pool, and increase material ductility. The deep energy penetration of the electron beam into the powder layer allows for the manufacturing of nearly fully dense materials with high mechanical strength, but it also produces rougher as-built surfaces with coarser feature resolutions compared to laser powder bed fusion. Such surface characteristics can increase contact resistance, induce additional pressure losses along the flow path, and alter flow behavior in bipolar plates. EBM systems are generally more complex, costly, and less accessible \cite{parenti2024techno, qian2015metal} than material extrusion or binder jetting additive manufacturing, especially for large-scale production.

Mo et al. \cite{mo2016additive} were the first that used EBM to manufacture a titanium-based GDL with an ordered microstructure. In comparison with conventional woven and sintered GDLs, EBM-fabricated layer delivered a much better electrochemical performance due to reduced ohmic losses enabled by its optimized microstructure and EBM high geometric fidelity. Figure~\ref{fig:12}h and i showed SEM images of the printed titanium GDL with a square pore size of 1.5 $\mathrm{mm}$ and a strut width of 500 $\mathrm{\upmu m}$. Through careful adjustment of processing parameters, a minimum wall thickness of 100 $\mathrm{\upmu m}$ was obtained. Although even smaller feature sizes were feasible, further refinement could introduce internal porosity within the struts. When implemented in PEMECs, the printed GDL yielded an 8\% increase in performance and efficiency at room temperature (see Figure~\ref{fig:12}j). In a preliminary study, Villemur et al. \cite{villemur2025comparison} looked into the EBM of Ti-6Al-4V for PEMFC bipolar plates. They printed coupons of porous BPs and subsequently coated them with titanium nitride. SEM analysis showed a high level of porosity, higher than 40\%, and the interfacial contact resistance below 10 $\mathrm{m\Omega \, cm^2}$ was measured for the TiN-coated coupons. There was no information on the dimensional accuracy, feature resolution, or surface roughness quantification of the prints. In addition, assessments of flow transport within the porous network or {{\it{in situ}}} electrochemical performance testing were not conducted. The as-printed coupons had a coarse surface texture, likely because of non-optimized EBM processing parameters. Another possible reason could be the partially fused or semi-sintered surrounding powder bed at high chamber temperatures, which tends to adhere onto the print surface. These partially sintered particles, together with the typically coarser powder feedstock used in EBM compared to laser-based processes, contributed to increased surface roughness and a more irregular surface morphology. 

While the studies above highlight the potential of the EBM process for BP fabrication, subsequent research on this topic has been quite limited. This gap may be due to the high cost of EBM and the rougher surface finish it produces compared to laser powder bed fusion.


\subsection{Directed energy deposition}
Directed energy deposition (DED) uses a concentrated energy source, such as an electron or laser beam, or electric arc, to liquefy material (e.g., wire or powder) as it is delivered onto the build area. While DED is primarily used to print metals and alloys, it can also process ceramics, metal matrix composites, and certain high-performance fiber-reinforced polymers. Powder feedstock allows for precise layer-wise material deposition with a resolution of $50-200$ $\mathrm{\upmu m}$, whereas wire deposition is typically coarser, around $500-1000$ $\mathrm{\upmu m}$ \cite{blakey2021metal}. Common DED processes include laser-based material deposition methods (including LENS and LMD), electron beam technologies used in freeform fabrication (EBF), as well as arc-driven wire additive manufacturing (WAAM), which differ in achievable resolution and deposition speed \cite{ahn2021directed, najmon2019review}. DED operates with larger melt pools and moderate cooling rates ($10^3-10^5$ $\mathrm{K/s}$), which often lead to a coarser microstructure compared to powder bed fusion additive manufacturing \cite{ettefagh2021corrosion}.

In a recent study, Cheng et al. \cite{cheng2024optimizing} reported the WAAM manufacturing of Ti-6Al-4V BP for proton exchange membrane water electrolysis (PEMWE). The corrosion performance of fabricated and heat-treated titanium plates was compared with that of conventionally manufactured plates in a simulated PEMWE environment. Small coupons of 5 mm $\times$ 10 mm $\times$ 2 mm were manufactured and sequentially polished, first mechanically and then with diamond and $\mathrm{SiO_2}$ suspension to relieve surface stresses. The microstructure, corrosion resistance, passive film characteristics, and electrochemical behavior were comprehensively characterized. The heat-treated coupons showcased the lowest corrosion current density of 54~$\mathrm{\upmu A /cm^2}$ and passive current density of 19.5~$\mathrm{\upmu A /cm^2}$. Similarly, Xu et al. \cite{xu2024microstructural} studied the microstructural evolution and corrosion behavior of LMD-processed nickel-based alloy for PEMFC bipolar plates thorough detailed microstructural characterization and electrochemical measurements. Fine alloy powder particles of size 7 to 53 $\mathrm{\upmu m}$ were deposited onto a 316L stainless steel substrate of 100 mm $\times$ 100 mm $\times$ 10 mm via LMD. The deposits were subsequently annealed at 1150~$^\circ$C and rapidly water-quenched for a consistent microstructure. The surfaces were mechanically polished and further finished by diamond powder. 

Despite these detailed material and electrochemical evaluations, neither study assessed the geometric fidelity or dimensional accuracy achievable with these DED processes. The dependence on extensive post-processing treatments, mechanical polishing followed by diamond and $\mathrm{SiO_2}$ suspension polishing, although effective in surface stress reduction, introduces additional processing complexity and may mask or alter the as-built geometry. Moreover, DED-based processes are primarily used for coating applications and repair purposes rather than for high-resolution fabrication of intricate geometries \cite{wang2007component, krishna2008functionally}. They are also susceptible to distortion and residual stress accumulation, particularly in thin-walled structures, due to high heat input and rapid deposition rates, which often necessitate substantial corrective post-processing interventions.

\subsection{Binder jetting and material jetting}
Binder jetting (BJ) is an additive manufacturing process wherein droplets of a liquid binder are selectively placed onto a thin spread of powder, commonly metal, ceramic, or composite materials. Wherever the binder is applied, material particles bond together and form a solidified layer \cite{kunchala20183d, mostafaei2021binder}. The green body is then cured and sintered to strengthen interparticle bonds and produce a dense final part. Binder jetting can achieve feature resolutions as fine as $20-50$ $\mathrm{\upmu m}$ \cite{neterebskaia2020inkjet}, which is desirable for fabrication of delicate PEMFC layers, including bipolar plates. Unlike binder jetting, material jetting (MJ) directly deposits droplets of photopolymers or polymers onto the build platform, rather than merely a binder. Each droplet solidifies instantly through UV curing or thermal processes. MJ is among the most precise additive manufacturing methods, with resolutions often below 50 $\mathrm{\upmu m}$.

Manogharan et al. \cite{manogharan2015binder} fabricated a complete SOFC via binder jetting. They sequentially printed the anode powder (nickel oxide–yttria stabilized zirconia), cathode powder (lanthanum strontium manganite 20\%), and ion-conducting electrolyte powder (yttria stabilized zirconia 8\%) to produce a 40 mm $\times$ 40 mm $\times$ 20 mm cell. Preliminary measurements quantified shrinkage during drying and sintering steps, followed by microstructural characterization and permeability analysis to evaluate porosity effects. They reported that with further optimization, cell porosity can be modified for optimal performance. Yoo et al. \cite{yoo20223d} also manufactured flexible, rubber-like BPs for bendable PEMFCs using material jetting of UV-curable photopolymers. Flow channels as small as 0.5 mm were precisely reproduced in parallel, serpentine, and pin-type layouts. {\it{In situ}} characterization and EIS measurements showed a peak power density of 87.1 $\mathrm{mW/cm^2}$ in the bent state, which was attributed to the compressive stress that improved interfacial contact and reduced ohmic and charge transfer resistances. Iqbal et al. \cite{iqbal2003corrosion} patented a binder jetting process of metallic BPs with precisely defined microchannels of 500 $\mathrm{\upmu m}$ width. To enhance surface conductivity and corrosion resistance, a conductive multilayer coating was applied onto the printed substrate. Subsequently, a thin overcoat of amorphous carbon or titanium suboxide was sputtered to infiltrate and seal fine-scale porosities within the underlying coating. In another embodiment, chemical anodization was performed to form a discontinuous aluminum oxide overlayer. This discontinuous morphology preserved electrical conduction routes and efficiently sealed pores. The corrosion behavior of the treated plates showed improved electrochemical stability and surface integrity that support long-term durability in fuel cell BP applications. Piri et al. \cite{piri20203d} also manufactured straight, parallel channels with a 0.4 mm $\times$ 0.4 mm square profile for PEMFC cathode BPs via both material jetting and SLA. SLA showcased excellent print resolution and closely reproduced fine channels. Ink jetting could replicate larger features of 0.635 mm with reasonable fidelity; however, when the channel size was reduced to 0.4 mm, residual support material and irregular channel profiles were observed. This suggests that ink jet printing is not able to reproduce features smaller than 0.5 mm. The authors also looked into the flow dynamics within the channels and assessed the pressure and velocity distributions. 

Binder jetting and material jetting bring high precision in the fabrication of BPs with smooth surface finishes. Binder jetting sequentially deposits a liquid binder to bond powders and is compatible with metals and ceramics, though curing and sintering post processing may introduce shrinkage and alter dimensions that mush be carefully managed. Material jetting, however, primarily processes photopolymer resins and, less commonly, metals; thus, conductive or protective coatings are typically required. Despite their fine resolution and multi-material capability compared with FFF and SLA additive manufacturing, both methods are still at an early stage of adoption for BP manufacturing and deserves further exploration.




\section{AM limitations, scalability, and industrial feasibility}
Conventional manufacturing methods have long been used for the fabrication of conventional PEMFC bipolar plates. As the second most expensive component after the catalyst layer, BPs contribute for roughly $20–30$\% of the total stack cost \cite{yang2019novel}, largely because of the high precision required for micro-channel fabrication with smooth surface finishes for which subtractive processes with material waste and multiple post-processing steps are inefficient. These limitations become even more pronounced for unconventional and non-intuitive flow channel geometries due to their insufficient resolution for fine features and reliance on multiple fabrication steps. This lack of design adaptability is a bottleneck in scaling up innovative BP designs from laboratory-scale demonstrations to industrial-scale production. Additive manufacturing, with its layer by layer fabrication, allows the realization of unconventional, multiscale, and multimaterial BPs with high geometric fidelity and often through monolithic fabrication. A growing body of literature has therefore studied additive manufacturing for BP and has improved the manufacturing readiness of PEMFC technologies beyond what is achievable through conventional processes. For instance, Trogadas et al. \cite{trogadas2018lung} printed lung-inspired fractal flow fields via DMLS of stainless steel, with the smallest channel profile of $200 \times 300$ $\mathrm{\upmu m}$ and well-defined channel geometry. Zhang et al. \cite{zhang2024heat} manufactured baffled flow fields via SLM with features as small as 0.3 mm and a dimensional accuracy of 0.05 mm. In another study \cite{zhang2023investigation}, they reported dimensional inaccuracy of less than 4\% in SLMed mesh and serpentine bipolar plates, with an average surface roughness of 0.135 $\mathrm{\mu m}$ after mild surface treatment. Munoz et al. \cite{munoz2023engineering} also studied SLA fabrication of bio-inspired interdigitated BPs with precise reproduction of sharp corners and very smooth surfaces. However, the commercialization and large-scale adoption of additive manufacturing for BPs remain limited by production costs and technological challenges \cite{wang2018techno, bar2002technical}. The limited resolution achievable by extrusion-based additive methods, such as FFF for polymer-based plates, and print-induced roughness with visible layer lines caused by the layer by layer filament deposition can affect cell performance by increasing ohmic resistance at the MEA interface and flow resistance inside the flow channels \cite{madadi2020influence}. Process-induced defects, poor interlayer adhesion, and pronounced mechanical anisotropy, which can lead to gas permeability and leakage, are also problematic \cite{jang2022effects, yang2019novel, madheswaran2024mwcnt, madheswaran2025pemfc}. Nevertheless, polymer-based FFF and binder jetting processes with lower material and energy requirements can reduce per-unit cost for small- to medium-scale production. Powder bed fusion methods, particularly SLM and DMLS/SLS processes, uses focused energy beams with small spot sizes of $20-100$ $\mathrm{\upmu m}$ and is able to precisely control the melt pools to create near fully dense prints with fine feature resolution. Thanks to this capability, PBF can be used for the fabrication of BPs with micorchannels and high surface quality. However, PBF techniques are energy-intensive, have slow build rates, and demand careful handling of metal powders \cite{parenti2024techno}, all of which contribute to high production costs. In addition, the steep thermal gradients and rapid solidification due to the high energy input can also induce residual stresses, often require post-processing heat treatments for stress relief, which further increases manufacturing costs. Conventional metal forming processes such as stamping and molding rely on dedicated tooling and tooling costs increase with the geometric complexity of the plate, and make these methods uneconomical for small-batch production, prototypes, or customized designs \cite{fitzlaff2025production}. Under such conditions, PBF-based additive manufacturing can become a competitive manufacturing route. BPs are conventionally made from graphite-based materials, either as pure graphite or as graphite-polymer composites \cite{song2020review}, due to their high conductivities and corrosion resistance. Nevertheless, graphite is brittle with poor processability through conventional methods such as CNC machining, and its thin thicknesses needed for high power density are fragile and hard to achieve \cite{yan2006performance, shimpalee2016investigation}. Alternatively, SLA and binder jetting allow for the creation of thin, intricate geometries of graphite bipolar plate. From a scalability perspective, additive manufacturing is particularly suitable for low-to-medium production volumes, as it eliminates the need for costly tooling, stamping dies, or molds typically required in conventional methods such as metal stamping, hydroforming, or injection/compression molding of graphite–polymer composites. This reduction in upfront capital investment allows rapid design iteration during development and accelerates innovation cycles. At large production volumes, however, conventional manufacturing generally becomes more cost-effective due to economies of scale. Established high-throughput processes, like stamping or molding, can produce thousands of plates per hour, whereas additive manufacturing layer-by-layer fabrication is comparatively slow with longer production times, higher machine utilization costs, and it relies on specialized feedstock materials. Despite these economic limitations for mass production, it enables highly efficient material usage by material deposition only where needed and reduces waste compared with subtractive methods. More importantly, its geometric freedom allows the creation of optimized BP designs that reduce mass transport limitations and deliver more homogeneous reactant concentrations over the catalyst layer. This can maximize the utilization of catalyst precious metals, such as Pt, and minimize local thermal gradients and the formation of hot or cold spots that can degrade catalyst lifespan. Thus, although the per-unit cost of additively manufactured BPs remains higher at large scales, the performance gains, particularly through improved catalyst efficiency, can partially offset fabrication costs.

\section{Process optimization and post-treatments in AM of bipolar plates}
Post-treatment in additive manufacturing refers to the additional processes applied to as-print materials at a raw, porous, or rough state to refine or modify their properties, dimensional fidelity, surface quality, or functionality. For PEMFC bipolar plates, post-treatment is particularly important as the plates must showcase a smooth surface finish to minimize interfacial contact resistance, high electrical conductivity and corrosion resistance, gas tightness, and accurately defined flow channel geometries. Since post-processing modifications often require additional time, energy, equipment or material, it is important to design the additive manufacturing process in a way that reduces the need for secondary treatments. In LPBF or DED processes, several surface finishing techniques are commonly used to remove partially fused particles and surface irregularities associated with AM fabrication. These techniques include polishing, abrasive flow machining, sandblasting, electropolishing, chemical etching, and laser surface modification \cite{alfieri2016reduction, gould2015performance, zhang2023investigation}. Such treatments help reduce surface roughness, improve electrical contact, and enhance flow characteristics within the channels. Heat treatment is another most common post-processing method for additively manufactured metals. It involves controlled heating and cooling cycles to relieve residual stresses generated by rapid solidification during printing, densify the printed material or alter its microstructure and mechanical properties \cite{karamimoghadam2022comparative, rajan2023heat, mahmood2022post, shiyas2021review}. For metals and metal matrix composites, common heat treatments include annealing to reduce residual stresses, hardening to enhance strength, and aging optimize microstructural properties \cite{karamimoghadam2022comparative, mahmood2022post, shiyas2021review}. Ceramics and ceramic matrix composites typically treated with annealing or stress-relief procedures, along with specialized sintering processes for densification and mechanical performance \cite{rahaman2017ceramic}. Surface modification and coating are also important post-processing steps in the fabrication of metallic bipolar plates. Although metals exhibit excellent mechanical strength and durability, they can suffer from corrosion and increased interfacial contact resistance when exposed to the acidic environment of PEMFCs. In response, protective and conductive coatings, such as carbon-based films, metal nitrides, or thin layers of noble metals are often deposited through physical vapor deposition, chemical vapor deposition, or electroplating \cite{celik2022development, trogadas2018lung, netwall2013decreasing, villemur2025comparison}. The additive manufacturing of BPs demands meticulous optimization of printing parameters to ensure a high quality print with higher infill density, dimensional fidelity, mechanical strength, and reliable electrochemical performance. Feedstock inconsistencies in powder-based processes can introduce inconsistencies in microstructure and material properties. For instance, differences in powder batches, such as variations in particle morphology and irregular particle shapes, agglomeration tendencies, and sensitivity to humidity, can compromise powder flowability and lead to uneven powder distribution across the powder bed. These inconsistencies often lead to uneven densification and structural defects within the print \cite{tan2017overview}. Therefore, control of powder production, handling, and storage conditions is essential to maintain process stability. Thermal management during printing is another critical parameter. The high localized temperatures generated by laser or electron beam sources can alter microstructure and introduce residual stresses if process parameters are not optimized. Adjustments to laser intensity and scanning velocity, and the thickness of each layer are necessary to control melt pool dynamics, minimize porosity, and reduce distortion and residual stresses. In PBF processes, partially melted particles and surface irregularities can also be addressed through meticulous process parameter optimization, combined with surface finishing or laser surface remelting. Recent advances in additive manufacturing process monitoring can further enhance production reliability. Adaptive process control and in-situ defect detection systems allow real-time identification of defects, melt pool instabilities, or dimensional inaccuracies, enable immediate adjustments during printing to achieve high-quality BPs \cite{wu2016detecting, ero2023optical}.

\section{Conclusion and outlook}
Despite the maturity of conventionally designed BPs and their adoption in practical PEMFC applications, they remain inadequate, as they typically derived from designer's limited intuition or incremental modifications of earlier designs with restricted and suboptimal performance. Extensive research efforts have therefore pursued the development of innovative BP designs that promote homogeneity of reactant distribution over catalyst sites to maximize catalyst utilization and prevent thermal gradients, lower pressure drop along the flow path to reduce the need for external blowers for reactant recirculation, and support efficient byproduct liquid water removal to prevent flooding, all of which contribute to reliable and efficient PEMFC operation. Such functional design solutions are typically non-intuitive with intricate geometries, and their realization and practical adoption depend on availability of advanced manufacturing routes that are both viable and industrially scalable. Additive manufacturing is capable of manufacturing these BPs that cannot be processed through conventional processing methods. This review article highlights the recent progress in application of additive manufacturing methods, including powder-based processes, extrusion-based methods, photopolymerization processes, and jetting approaches, for BP fabrication. The current performance of additively manufactured bipolar plates, associated problems and limitations, and possible solutions for improvement have also been critically discussed for each process. Among these processes, SLA and DLP from vat photopolymerization family demonstrate strong potential for polymeric BPs with intricate microchannel patterns and microporous flow fields with smooth surface finishes, high print resolution as small as 20 $\mathrm{\upmu m}$, and minimal post-treatments. However, they are less common PEMFC BP fabrication due to their incompatibility with metallic materials and high-performance thermoplastics such as PEEK, PEKK. Inclusion of conductive and reinforcement fillers or conductive surface coatings could ensure adequate chemical durability, electrical conductivity, and mechanical properties, and thereby enhance the applicability of vat photopolymerization. Powder bed fusion processes typically use fine powder particles and can create very thin layers and smooth as-built surfaces. They are capable of producing dense components with nearly isotropic mechanical properties comparable to those of conventionally manufactured materials, which make this method one of the most promising AM approaches for high-performance metallic bipolar plates. Nevertheless, the relatively high energy consumption, potential residual stress development, and overall production cost remain important challenges. Continued optimization of processing parameters and improvements in process efficiency are therefore, necessary to enable broader industrial adoption. DED-based processes, on the other hand, generally involve larger melt pools and moderate cooling rates, which often lead to coarser microstructure and residual stress accumulation, particularly in a thin-walled plate. Substantial post-treatment may be required to achieve the desired dimensional accuracy and surface quality. For this reason, DED techniques are primarily used for coating applications and repair purposes and are less recommended for high-resolution bipolar plates. Attempts have also been made to use FFF for bipolar plate, but is less precise than many other additive methods. Their surface roughness, visible layer lines, and dimensional inaccuracies caused by warpage may disrupt reactant transport and induce leakage. The low electrical conductivity of polymers and presence of interlayer voids and open porosity are also problematic. Although post-processing approaches such as laser surface treatment, infiltration, and conductive coatings can partially address these issues, they increase manufacturing complexity and production time, which are not recommended.

Taken together, the findings from the literature suggest that additive manufacturing has strong potential as a manufacturing route for the next generation of PEMFC bipolar plates. The capability of AM technologies to produce highly customized geometries, reduce material waste, and enable integrated functional designs presents a pathway toward more efficient and sustainable fuel cell component manufacturing. Future research should focus on improving material conductivity, optimizing process parameters for higher dimensional accuracy and surface quality, developing hybrid or multimaterial printing strategies, and reducing production costs to facilitate the large-scale adoption of additively manufactured BPs in commercial PEMFC systems.


\section*{Acknowledgements}
This work was supported by the Natural Sciences and Engineering Research Council of Canada [NSERC under grant RGPIN-217525].

\section*{CRediT authorship contribution statement}
Zahra Kazemi: Conceptualization, Formal analysis, Visualization, Writing – original draft. Kamran Behdinan: Conceptualization, Funding acquisition, Project administration, Supervision, Writing – review \& editing.

\section*{Declaration of generative AI and AI-assisted tools in the writing process}
During the preparation of this manuscript, the authors used ChatGPT to improve language clarity and readability. All content generated with the assistance of this tool was subsequently reviewed and edited by the authors, and they take full responsibility for the manuscript content.

\section*{Declaration of competing interest}
The authors declare that they have no known competing financial interests or personal relationships that could have appeared to influence the work reported in this manuscript.  

\bibliographystyle{unsrt} 
\bibliography{example}


\end{document}